%% file: 00-paper.tex
\documentclass[nonacm]{acmart}

\usepackage[backgroundcolor=red,textsize=footnotesize]{todonotes}   
\usepackage[utf8]{inputenc} 
\usepackage[T1]{fontenc}    
\usepackage{hyperref}       
\usepackage{url}            
\usepackage{booktabs}       
\usepackage{amsfonts}       
\usepackage{nicefrac}       
\usepackage{microtype}      
\usepackage{graphicx}
\usepackage{float}
\usepackage{adjustbox}
\usepackage{multirow}		
\usepackage[para,online,flushleft]{threeparttable}
\usepackage{array}
\usepackage{bm}
\AtBeginDocument{%
  \providecommand\BibTeX{{%
    \normalfont B\kern-0.5em{\scshape i\kern-0.25em b}\kern-0.8em\TeX}}}

\RequirePackage[normalem]{ulem} \RequirePackage{color}\definecolor{RED}{rgb}{1,0,0}\definecolor{BLUE}{rgb}{0,0,1} 

\setcopyright{none}
\copyrightyear{2020}
\acmYear{2020}
\acmDOI{}

\acmJournal{JACM}
\acmVolume{xx}
\acmNumber{xx}
\acmArticle{xx}
\acmMonth{8}

\makeatletter
\def\blfootnote{\xdef\@thefnmark{}\@footnotetext}
\makeatother

\begin{document}

\title{A Simple Model for Portable and Fast Prediction of Execution Time and Power Consumption of GPU Kernels}

%
\author{Lorenz Braun}
\email{lorenz.braun@ziti.uni-heidelberg.de}
\affiliation{%
  \institution{Institute of Computer Engineering, Heidelberg University}
  \country{Germany}  
}

\author{Sotirios Nikas}
\email{sotirios.nikas@uni-heidelberg.de}
\affiliation{%
  \institution{Engineering Mathematics and Computing Lab, Heidelberg University}
  \country{Germany}  
}

\author{Chen Song}
\email{chen.song@iwr.uni-heidelberg.de}
\affiliation{%
  \institution{Engineering Mathematics and Computing Lab, Heidelberg University}
  \country{Germany}  
}

\author{Vincent Heuveline}
\email{vincent.heuveline@uni-heidelberg.de}
\affiliation{%
  \institution{Engineering Mathematics and Computing Lab, Heidelberg University}
  \country{Germany}  
}

\author{Holger Fröning}
\email{holger.froening@ziti.uni-heidelberg.de}
\affiliation{%
  \institution{Institute of Computer Engineering, Heidelberg University}
  \country{Germany}  
}

\renewcommand{\shortauthors}{Lorenz Braun, et al.}

\begin{abstract}
Characterizing compute kernel execution behavior on GPUs for efficient task scheduling is a non-trivial task.
We address this with a simple model enabling portable and fast predictions among different GPUs using only hardware-independent features.
This model is built based on random forests using 189 individual compute kernels from benchmarks such as Parboil, Rodinia, Polybench-GPU and SHOC.
Evaluation of the model performance using cross-validation yields a median Mean Average Percentage Error (MAPE) of 8.86--52.0\% for time and 1.84--2.94\% for power prediction across five different GPUs, while latency for a single prediction varies between 15 and 108 milliseconds.
\end{abstract}



\maketitle

\input{01-introduction} 
\input{02-background} 
\input{03-methodology} 
\input{04-groundtruth}

\input{06-modelling} 
\input{07-portability} 
\input{08-discussion} 
\input{09-summary} 

\begin{acks}
	This work is supported in part by the Federal Ministry of Education and Research of
	Germany in the framework of Mekong project (FKZ: 01IH16007). The authors would like to thank Ullrich Koethe at Heidelberg University and Kai Polsterer at Heidelberg Institute for Theoretical Studies for their help on machine learning methods and models.
	We furthermore acknowledge the helpful comments of the reviewers throughout the publication process.
\end{acks}

\bibliographystyle{ACM-Reference-Format}
\bibliography{ref}


\end{document}

%% file: 01-introduction.tex
\section{Introduction}
\blfootnote{}


GPUs are massively parallel multi-processors, and offer a tremendous amount of performance in terms of operations per second, memory bandwidth and energy efficiency. As a result, they are being used pervasively in areas beside visual computing, including scientific and technical computing, machine learning and data analytics.
Programs running on GPUs are expressed in compute kernels, which are code regions compiled separately for such co-processors, but called from the main host processor. 
As the GPU execution model demands for a high amount of structured parallelism, such kernels are typically well-structured and behave regularly for avoiding fine-grained control flow.

GPU computing is a prime example for heterogeneous computing, which ultimately 
requires tools reasoning about the most suitable processors for a given workload 
respectively kernel.
Heterogeneity adds complexity which can be tackled by schedulers, which automatically reason about task placement.
In this regard, predictive modeling can assist the scheduler,
as predictions of execution time allow to select the 
fastest processor for a given workload.
Similarly, for instance when partitioning a task for multi-GPU execution, execution time predictions can help to avoid too small work items, or to ensure enough overlap in between compute and communication tasks.
Such predictions also apply for other tasks including system provisioning and procurement,
as well as replacing execution time by power consumption as key metric.

If there exists such a predictive model (in the following: \emph{model}), 
which is based solely on hardware-independent features, it would allow to reason 
about time and power for different GPU architectures and models, enabling to 
identify the most effective ones in terms of performance per unit cost.
Such features can include instruction counts (floating-point operations, integer operations, memory operations on 
different address spaces, etc.), or the thread hierarchy of the kernel in execution (kernel launch configuration), but not hardware-dependent features like cache hit rates.


Various performance and power models already exist for GPUs
\cite{AGPU,MadougoulandscapeGPGPUperformance2016,WuGPGPUperformancepower2015,CarrollImprovedAbstractGPU2017,Zhangquantitativeperformanceanalysis2011,HongAnalyticalGPUModel2009,HongintegratedGPUpower2010,HuangGPUMechGPUPerformance2014,Baghsorkhiadaptiveperformancemodeling2010,SongSimplifiedAccurateModel2013,ChenStatisticalGPUpower2011,Lim2014,5598315, 8327055, 7920860, 8675207, Johnston_2018, 7778637, Lehnert:2016:PPR:2990973.2991003, SalariaDPM19, 6468530, Reisert, ArafaScalableGPU2019, GuerreiroGPUStaticModeling2019}.
They are usually based on: (1) executing the program under observation, with additional costs depending on the required execution statistics; (2) collecting execution statistics using the processor's performance counters; and (3) inferring executing time and power consumption based on these statistics. 
As a result, such models rely on a variety of input features, in particular also hardware-related features like cache hit rates.
Notably, some models yield good prediction performance without the use of such hardware-related features \cite{HongintegratedGPUpower2010,Baghsorkhiadaptiveperformancemodeling2010}.
Some previous work has used analytical models (e.g., \cite{5598315,HongAnalyticalGPUModel2009,HongintegratedGPUpower2010, ArafaScalableGPU2019}), but machine learning based methods like for instance Artificial Neural Networks (ANNs) have demonstrated a highly improved accuracy (e.g., \cite{SongSimplifiedAccurateModel2013, WuGPGPUperformancepower2015, GuerreiroGPUStaticModeling2019}).
Concerning vendor tools, RAPL by Intel and NVIDIA's NVML are power measurement tools that are sometimes based on modeling techniques, and, in particular, require to execute a program in order to obtain knowledge about power consumption.
While various solutions have been proposed, two particular downsides are apparent: first, it is not documented how well those models fit to other GPU architectures (lack of portability). Second, there are only few publicly available performance and power models for GPUs~\cite{ArafaScalableGPU2019, GuerreiroGPUStaticModeling2019}
- however those are based on static analysis of kernel assembly, and as such limited to workloads with rather similar control flow behavior among different threads (lack of availability).
GPU performance highly depends on the presence of structured parallelism, with corresponding performance bugs including branch divergence, memory coalescing issues, and shared memory bank conflicts. 
Furthermore, shared memory as an explicitly controlled element of the memory hierarchy requires applications to be locality-optimized, including a maximization of data re-use. 
Last, GPUs are prime examples of the Bulk-Synchronous Parallel (BSP) execution model \cite{valiant1990}, which demands for latency tolerance by a large amount of parallel slackness, such that way more threads are in execution than execution units present.
As a result, good GPU code is usually well-structured, locality-optimized and latency-tolerant, which is inline with our experience and review of publicly available GPU benchmark 
suites \cite{CheRodiniabenchmarksuite2009,StrattonParboilRevisedBenchmark,
  DanalisScalableHeterogeneousComputing2010,
  Grauer-GrayAutotuninghighlevellanguage2012}.
If this holds true, we hypothesize that the importance of hardware-dynamic effects with regard to execution time and power consumption is limited, such that predicting those based solely on hardware-independent features is feasible.
In detail, GPU kernel behavior should be mainly determined by static code features such as instruction counts and kernel launch configuration, and static hardware parameters like operating frequency, cache size, access latency, number of execution units, and general architecture.
Thus, a model trained on a particular GPU architecture, thereby capturing the static hardware parameters, should be able to accurately predict kernel behavior based solely on static code features.
Note that this does not suggest that for instance caches are of no relevance, instead we argue that well-behaving GPU applications will ensure that caches are effectively used, such that their influence on execution time is limited.
In this regard, an adversarially-written application will surely result in a large influence and a contradicted hypothesis.

We therefore propose a method and model for predicting kernel execution time and power consumption based on machine learning techniques, which is:

\begin{itemize}
	\item \textbf{Simple}: it is based on features that can be derived quickly and with minimal overhead in terms of additional execution time due to instrumentation.
	\item \textbf{Portable}: it can be easily ported to other GPU architectures 
	by simply retraining the model, based on the same feature selection and general methodology.
	\item \textbf{Fast}: as it is based on a simple random forest model, no large amount of computation is required to produce a prediction.
\end{itemize}

As a result, (1) it requires only minimal overhead for profiling (model feature acquisition), 
(2) it allows for provisioning tasks as it can be easily ported to a variety of different GPU types, 
and (3) it is suitable for a use in schedulers, which usually require that the time for scheduling decisions is orders of magnitude shorter than the execution of the program.

The detailed contributions are as follows:
\begin{itemize}
  \item A portable profiling infrastructure for acquisition of input and output features, used for training and possible re-training for portability reasons.
  \item A model suitable for a small input feature set, which is fast and 
  sufficiently accurate
  for runtime decisions on scheduling (heterogeneity) and orchestration of kernels and data movements (prefetching respectively overlap).
  \item An evaluation of method and model demonstrating prediction performance, 
  prediction speed, and prediction portability: for a variety of GPU kernels 
  from various benchmark suites, predictions for five different GPU types are 
  evaluated (NVIDIA K20, GTX1650, Titan Xp, P100 and V100).
  \item Method, model, measurement infrastructure and training tool are made publicly available \footnote{\url{https://github.com/UniHD-CEG/gpu-mangrove}}.
\end{itemize}

%% file: 02-background.tex
\section{Background}

In the following, we will shortly review GPUs and CUDA, random forests as fundamental machine learning method, and related work in the context of predictive modeling.

\subsection{GPU architecture and programming}

The following introduction of GPUs is based on CUDA nomenclature, even though OpenCL is very similar except for different naming.

GPUs are massively parallel processors, executing multiple thousands of light-weight threads formed into a hierarchy: 
multiple threads are grouped into thread blocks\footnote{Also referred to as Cooperative Thread Array (CTA).}, with the possibility of fast barrier synchronization and data exchange using shared memory structures equally fast as conventional caches.
Multiple blocks form a thread grid, which is specified as part of the kernel launch configuration.
Thus, a thread grid is a kernel in execution.
For pre-Volta GPUs, synchronization among different thread blocks is not supported, as GPUs miss strong progress guarantees due to a lack of preemption.
GPUs do not execute single threads individually, instead multiple threads (typically 32) form a thread warp which is the main unit for scheduling.
As a result, all threads of a warp share a single instruction stream, and non-coherent control flow in a warp results in serialization.

While such GPUs only have been able to support lock-free algorithms, Volta GPUs introduced an independent thread scheduling, which supports starvation-free algorithms, such as mutual exclusion, even in the presence of warp divergence \cite{Choquette2018Volta}.
This independent progress is based on the compiler identifying visible execution steps, such as a barrier or an atomic operation.
Notably, independent thread scheduling as publicly described does not explicitly preclude warp-based execution, albeit a dynamic re-formulation of warps is most likely employed.

The memory hierarchy of a GPU is flat and thus very different from general-purpose processors like CPUs.
Threads can operate on register space as private memory, while thread blocks can make use of shared memory as cache-like memory resource.
The main memory resource of a GPU is on-card GDDR-based high-throughput memory, called global memory or device memory.
Unlike registers and shared memory, the lifetime of global memory exceeds the lifetime of a single kernel.
Also, global memory is the main resource for interactions between host and GPU.

There also exist caches on a GPU, but as a GPU relies on latency tolerance and not latency minimization, caches can be small.
In particular, unlike a CPU, a GPU does not make use of caches to reduce average (global) memory access latency, instead its main purpose is to reduce contention on lower levels of the memory hierarchy.
For latency tolerance, GPUs are prime examples for the Bulk-Synchronous Parallel (BSP) execution model \cite{valiant1990}, which requires a large amount of parallel slackness in the form of orders of magnitude more threads in execution than physical processing units present.

Still, GPUs consist of up to thousands of processing units, which are grouped into so-called Streaming Multi-Processors (SMs).
A thread block can execute only on a single SM, and, as a result, there is no interaction among different SMs except for global memory.
Hence, GPU architectures efficiently scale with the number of SMs, and kernels written once hopefully observe excellent performance portability on more recent GPUs.

With regard to the present work, notice in particular that common code optimization techniques for GPUs require that code is well-structured and behaving regularly with regard to coherent control flow and thread behavior.
Otherwise, multiple performance penalties exist: unstructured access to shared memory might result in bank conflicts and access serialization. 
Similarly, unstructured access to global memory results in non-coalesced accesses to off-chip DRAM modules.
Thread-individual control flow usually causes branch divergence penalties, as instructions are shared at warp level and non-coherent branching is handled by collectively execution all paths with appropriate masking of results.

%


\subsection{Random Forests}

Random forests are a machine learning method based on ensemble learning for 
either regression or classification tasks \cite{BreimanLeoRandomforests2001}. 
During training, multiple decision trees are constructed.
On each node an input feature will be compared to a threshold and the result determines the next 
node to be processed until a leaf with an output value is reached. 

Construction of a tree is controlled by multiple parameters. In the case of 
the scikit-learn implementation \cite{scikit-learn}, the main parameters to 
adjust are the number of estimators (trees) \emph{n\_estimators}, the maximum depth of the trees \emph{max\_depth} and 
\emph{max\_features} as the number of features being used when splitting 
a node in a tree. More estimators typically lead to better result, but  
take more time to train and to predict. Low max\_features 
parameters reduce variance but increase bias. 
Last, there are different variations of the split criterion, which measures the quality of a split.


Random forest algorithms allow to compute relative feature importance by analyzing the relative rank, which is the depth of a decision node in a tree respectively the feature used for that decision node. 
These importances can be used to check whether the trained model behaves as expected. 

\subsection{Related Work}

\begin{table}
	\scriptsize
	\centering
	\renewcommand{\arraystretch}{1.1}
	\begin{tabular}{m{0.04\textwidth} | m{0.011\textwidth} | m{0.011\textwidth} | >{\raggedright}m{0.062\textwidth} | >{\raggedright}m{0.2\textwidth} | >{\raggedright}m{0.18\textwidth} | >{\raggedright}m{0.11\textwidth} | m{0.11\textwidth} | m{0.035\textwidth} }
			
		Source & T & P & \centering{Model} & \centering{Accuracy} & \centering{Portability} & \centering{Input source} & \centering{Dataset} & DVFS \\
		\hline
		\hline
		\cite{HongAnalyticalGPUModel2009} & $\checkmark$ &  & AM & GMAE: 5.4-13.3\%  & 4 NVIDIA GPUs (Fermi) & NVIDIA PTX, custom & 20 apps &   \\
		\hline
		\cite{Baghsorkhiadaptiveperformancemodeling2010}  & $\checkmark$ &  & AM & good agreement between predicted and observed & 1 NVIDIA GPU (Tesla)  & PDG & 4 apps & \\
		\hline
		\cite{5463058} & $\checkmark$ &  & RM, ANN & median error: 1.16-6.65\%  & 1 CPU & Xen-specific & 4 apps & \\
		\hline
		\cite{HongintegratedGPUpower2010} & $\checkmark$ & $\checkmark$ & EM, AM & GME: 2.7\% (micro-benchs), 8.94\% (merge) & 1 NVIDIA GPU (Tesla)  & GPUOcelot & 20 apps &\\
		\hline
		\cite{5598315} & & $\checkmark$  & LR & ASE: 54.9\%, AER: 4.7\% & 1 NVIDIA GPU (Fermi)  & CUDA Profiler & 49 kernels & \\
		\hline
		\cite{Zhangquantitativeperformanceanalysis2011} & $\checkmark$ &  & TM & error: 5-15\% & 1 NVIDIA GPU (Fermi) & Barra, cubin, nvcc, HW res & 3 apps, micro-benchmarks & \\
		\hline	
		\cite{ChenStatisticalGPUpower2011}  &  & $\checkmark$ & RF, RT, LR &APE: 7.77\%, 11.68\%, 11.7\% & 1 NVIDIA GPU (Tesla)  & GPGPUSim & 52 kernels & \\
		\hline
		\cite{6468530} & $\checkmark$ &  & AM & ARE plots & Intel CPU \& NVIDIA GPU (Fermi) & Aspen & 4 kernels &\\
		\hline
		\cite{SongSimplifiedAccurateModel2013} & $\checkmark$ & $\checkmark$ & MLR, ANN & AAPER: 6.7\% (T), 2.1\% (P) & 2 NVIDIA GPUs (Fermi) & CUPTI, custom & 20 kernels &\\ 
		\hline
		\cite{Lim2014}  & & $\checkmark$ & EM & AE: 7.7\% (micro-bench), 12.8\% (merge)  & 1 NVIDIA GPU (Fermi) &  MacSim, DRAMSIM & 23 apps &\\
		\hline
		\cite{HuangGPUMechGPUPerformance2014}  & $\checkmark$ &  & AM, IA & AE: 13.2\% (RR), 14.0\% (GTO) & Fermi-like architecture & GPUOcelot & 40 kernels & \\
		\hline
		\cite{WuGPGPUperformancepower2015} & $\checkmark$ & $\checkmark$ & ANN & AE: 15\% (T), 10\% (P) & 6 AMD GPUs (GCN) & AMD CodeXL & 108 kernels & $\checkmark$ \\
		\hline
		\cite{Lehnert:2016:PPR:2990973.2991003} & $\checkmark$ &  &  LR, KNN &median divergence: 10\% & 2 NVIDIA GPUs (Fermi, Kepler) & custom & 1 app (SpMV) &\\  
		\hline
		\cite{7778637} & $\checkmark$ &   & AM, LR, SVM, RF  & MPA (pred/meas): 0.75-1.5\% (ML), 0.8-1.2\% (AM) & 9 NVIDIA GPUs (Kepler, Maxwell) & nvprof, custom & 9 apps &\\
		\hline
		\cite{7920860} & $\checkmark$ & $\checkmark$ & RF & MAPE: 25\% (T), 12\% (P) & AMD CPU+APU & AMD CodeXL & 73 apps & $\checkmark$\\
		\hline
		\cite{Reisert} & $\checkmark$ &  & EM & SMAPE: 12.97\% & CPU-GPU (no info) & Score-P & 7 apps \\ 
		\hline
		\cite{CarrollImprovedAbstractGPU2017}  & $\checkmark$ &  & AM & predicted/observed: 1.5\% (vector ops), 0.76\% (matrix ops), 5.49\% (reduction)  & 1 NVIDIA GPU (Kepler)  & custom & 3 apps & \\
		\hline	
		\cite{WangGPGPUPerformanceEstimation2018} & $\checkmark$ &  & AM & MAPE: 3.5\% & 1 NVIDIA GPU (Maxwell) & Nsight, custom, hard spec & 12 kernels & $\checkmark$\\
		\hline
		\cite{8327055} &  & $\checkmark$ & SR & MAE: 7$\%$ (Pascal), 6$\%$ (Maxwel), 12$\%$ (Kepler) & 3 NVIDIA GPUs (Pascal, Maxwell, Kepler)  & CUPTI, custom & 83 apps & $\checkmark$\\
		\hline
		\cite{Johnston_2018} & $\checkmark$ &   & RF & MAE: 1.2\% & 3 CPUs, 1 Xeon Phi, 5 NVIDIA GPUs (Kepler, Pascal), 6 AMD GPUs & AIWC & 37 kernels &\\
		\hline
		\cite{ArafaScalableGPU2019} & $\checkmark$ & & HM & w/in 10\% to real device performance & 2 NVIDIA GPUs (Maxwell, Kepler) & PTX, NVIDIA Visual Profiler & 10 kernels &\\
		\hline
		\cite{8675207} & $\checkmark$ &   & HM & MAPE: 17.04\% & 2 Kepler GPUs, 2 NVIDIA GPUs (Maxwell) & LLVM, custom & 20 kernels &\\
		\hline
		\cite{SalariaDPM19} & $\checkmark$ & & AM & AE: 9.4\%  & 7 NVIDIA GPUs (Kepler, Maxwell, Pascal, Volta)  & no information & 30 apps &\\
		\hline
		\cite{FanPredictableGPUs2019} & $\checkmark$ & $\checkmark$ & OLS, PR, SVR & RMSE: 6.68-11.13\% (speedup), 5.65-15.10\% (energy) & 1 NVIDIA GPU (Maxwell) & LLVM & 118 kernels & $\checkmark$\\
		\hline
		\cite{GuerreiroGPUStaticModeling2019} & $\checkmark$ & $\checkmark$ & LSTM & MAE: 5.35-7.85\% (P), 9.9-19.3\% (T) & 4 NVIDIA GPUs (Turing, Volta, Pascal, Maxwell) & PTX & 169 kernels & $\checkmark$\\
		\hline
		\hline
		Ours  & $\checkmark$ & $\checkmark$  & RF & MAPE: 8.86-52.0\% (T), 1.84-2.94\% (P)  & 5 NVIDIA GPUs (Kepler, Pascal, Volta, Turing) & CUDA Flux &189 kernels (T), 168 kernels (P) &
	\end{tabular}
	\renewcommand{\arraystretch}{1.0}
	
	\caption{An overview of related work, showing prediction target (time [T], power [P]), used model, accuracy, portability, input feature source, and dataset size.}
	
	\label{tab:related_work}
	
\end{table}

In last years, performance and power modeling of GPUs are attracting considerable interest and several approaches have been proposed.
A summary of related works sorted in chronological order including prediction (execution time or power consumption or both), model, accuracy, portability, input source, dataset size and support for different dynamic voltage and frequency scaling (DVFS) settings is given in Table~\ref{tab:related_work}.

The most common approach is using machine learning methods such as random forest (RF) \cite{7920860, Johnston_2018, 7778637}, support vector machines (SVM) \cite{7778637}, artificial neural networks (ANN)  \cite{SongSimplifiedAccurateModel2013,WuGPGPUperformancepower2015, 5463058}, long short-term memory networks (LSTM) \cite{GuerreiroGPUStaticModeling2019}, k-nearest-neighbor (KNN) \cite{Lehnert:2016:PPR:2990973.2991003}, and so forth.  
The other common approach is using regression-based models such as statistical regression (SR) \cite{8327055}, regression model (RM) \cite{Barnes:2008:RAS:1375527.1375580}, regression trees (RT) \cite{ChenStatisticalGPUpower2011}, linear regression (LR) \cite{5598315}, multiple linear regression (MLR) \cite{SongSimplifiedAccurateModel2013} , ordinary least squares linear regression (OLS), LASSO, polynomial regression (PR) and support vector regression (SVR)  \cite{FanPredictableGPUs2019}. 
Machine learning and regression methods provide an accurate prediction, albeit, tedious effort is required on feature engineering. 
However, this fundamental issue can be overcome by automatic methods evaluating the feature impact on the accuracy of the model. 

The other major approach is to use analytical models (AM). 
One example is Aspen as a domain specific language for analytical performance modeling \cite{6468530}, which basically requires to rewrite an application in this language.
Another analytical model considers the number of running threads and memory bandwidth for predicting performance \cite{7778637}. 
There is also analytical model using the novel collaborating filtering based modeling technique to predict the performance \cite{SalariaDPM19}.
Alternatively, the interval analysis (IA), which differs from traditional analytical models, uses both trace-driven functional simulators and analytical model to estimate core-level performance \cite{HuangGPUMechGPUPerformance2014}.
In this approach, GPUMech, an interval analysis-based performance modeling technique for GPU architectures was used for modeling two popular warp scheduling policies, namely round-robin scheduling (RR) and greedy-then-oldest (GTO), respectively.

Furthermore, there are approaches combining the aforementioned methods, leading to a hybrid model (HM) \cite{8675207}. 
For example, combing an analytical model and with an event-based simulation of the code \cite{ArafaScalableGPU2019},
or models as throughput model (TM) \cite{Zhangquantitativeperformanceanalysis2011} and empirical model (EPM) \cite{Reisert, HongintegratedGPUpower2010, Lim2014} exist as well.

Numerous metrics have been used for measuring the accuracy of models such as
Average Absolute Prediction Error Rate (AAPER) \cite{SongSimplifiedAccurateModel2013}, Average Error (AE) \cite{Lim2014, HuangGPUMechGPUPerformance2014, WuGPGPUperformancepower2015, SalariaDPM19}, Average Error Ratio (AER) \cite{5598315},
Average Percentage Error (APE) \cite{ChenStatisticalGPUpower2011}, Absolute and Relative Error (ARE) \cite{6468530}, 
Average Squared Error (ASE) \cite{5598315}, Geometric Mean of Absolute Error (GMAE) \cite{HongAnalyticalGPUModel2009}, Geometric Mean of the Error (GME) \cite{HongintegratedGPUpower2010}, Mean Absolute Error (MAE) \cite{8327055, GuerreiroGPUStaticModeling2019, Johnston_2018}, Mean Absolute Percentage Error (MAPE) \cite{7920860, WangGPGPUPerformanceEstimation2018, 8675207}, Mean Prediction Accuracy (MPA) \cite{7778637}, Mean Squared Error (MSE) \cite{7778637}, Root Mean Square Error (RMSE) \cite{FanPredictableGPUs2019} or Symmetric Absolute Percentage Error (SMAPE) \cite{Reisert}. Different performance metrics are used as they serve different purposes \cite{Botchkarev2018PerformanceM}, but a detailed explanation is beyond the scope of this work.

Besides accuracy, another important characteristic of a model is to enable portability across different GPUs or other accelerators. 
Several studies \cite{8327055,SongSimplifiedAccurateModel2013,7920860,WuGPGPUperformancepower2015,HongAnalyticalGPUModel2009, Johnston_2018,7778637,ArafaScalableGPU2019,8675207,GuerreiroGPUStaticModeling2019,Lehnert:2016:PPR:2990973.2991003,SalariaDPM19} have been conducted on this direction, while many other works focus on one single processor.


Most often, the input features for training the model are performance counters, which are required from tools including CUPTI, AMD CodeXL, nvprof, LLVM, Score-P, nvcc, Barra simulator, cubin generator, GPGPUSim, MacSim, DRAMSIM, GPUOcelot (PTX emulator), Architecture Independent
Workload Characterization (AIWC), among others.
However, most of the studies do not rely exclusively on those tools but also develop on their own using custom microbenchmarks, further code analysis, kernel compilation information, hardware specifications, analytical equations, program dependence graphs (PDG) and others.

As previously mentioned, a representative training dataset is important for the model's generalization capability.
The size of such a dataset varies highly among the studies, ranging from one single application to up to 169 different kernels.
Note that it often remains unclear in the references if an application consists of multiple kernels which are treated independently or not.

More recently, there is the tendency on predicting performance and power consumption for different DVFS settings, namely for different memory and core frequencies. Subsequently, those works aim to determine the best DVFS settings providing the best performance with the minimum power consumption to leverage energy efficiency \cite{7920860, WangGPGPUPerformanceEstimation2018, 8327055, FanPredictableGPUs2019, GuerreiroGPUStaticModeling2019}.

Our work distinguishes from most of these related works by using only hardware-independent input features for model training.
Only quite few related works are also based on static input features \cite{HongintegratedGPUpower2010,GuerreiroGPUStaticModeling2019,FanPredictableGPUs2019,Baghsorkhiadaptiveperformancemodeling2010}, while the vast majority requires a comprehensive application analysis prior to prediction.
Last, we are aware of only two other works being publicly available \cite{ArafaScalableGPU2019, GuerreiroGPUStaticModeling2019}, and will discuss them in Section~\ref{sec:discussion}.


%% file: 03-methodology.tex
\section{Methodology}
\begin{figure}[!ht]
	\includegraphics[width=0.48\linewidth]{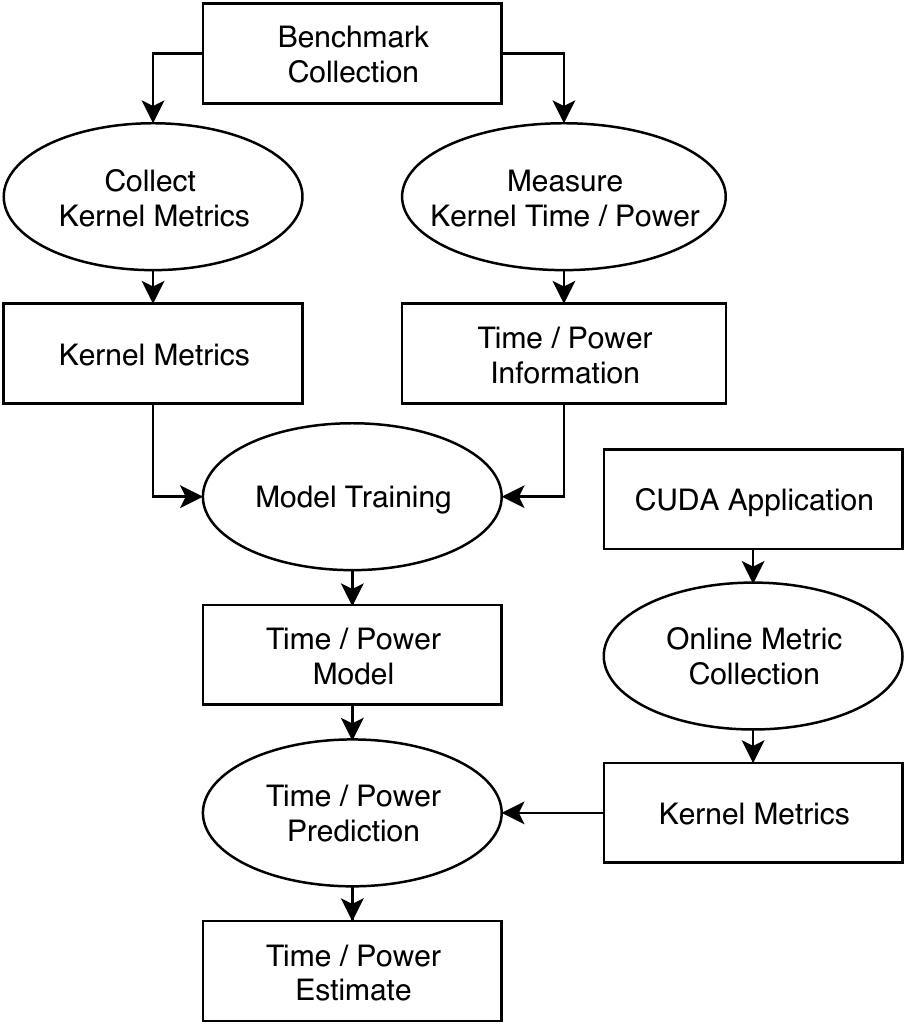}	
	\caption{Workflow for execution time and power prediction using CUDA Flux. 
	Rectangular nodes represent data and oval nodes processes.}
	\Description{Workflow for execution time and power prediction using CUDA Flux. 
		Rectangular nodes represent data and oval nodes processes.}
	\label{fig:workflow}
\end{figure}

As machine learning methods have been proved to be highly accurate for modeling and predicting performance of processors \cite{SongSimplifiedAccurateModel2013,WuGPGPUperformancepower2015}, we are also relying on such techniques.
Figure~\ref{fig:workflow} provides a summary of the workflow. 
The left part mainly covers the training part, based on collecting metrics as input features,  execution time and power consumption as ground truth.
Thus, samples are formed of a space $X$ of input feature vectors $\bm{x_i}$, each with a label $y_i$, all labels forming the output space $\bm{Y}$.
$y_i$ are the labels for training, respectively the target values for inference.
Generally spoken, the goal of training procedure is to find a model or function $g: X \rightarrow Y$ for which a scoring function $f: X \times Y \rightarrow \mathbb{R}$ is maximized, in other words, the error of prediction is minimized.


The inference process is shown on the right of Figure~\ref{fig:workflow}.
Collected metrics from CUDA applications will be used by trained model $g$ to predict execution time or power consumption.
In our study, we use a collection of four benchmark suites in order to obtain a broad dataset for execution time and power consumption, respectively.

Commonly used metrics for the scoring function include 
Mean Absolute Error (MAE), Mean Squared Error (MSE) and R-squared Error ($R^2$). 
Execution time measurements have shown that kernels last from a few microseconds to 
multiple seconds. 
It implies that, if the short kernels have too few contributions to the scoring function $f$, they 
are inevitably to be considered as noise.
Absolute-value-based errors, e.g. MAE and MSE, 
are not a good fit for our dataset, 
because the errors in long-running kernels are weighted more than short ones.
Therefore, a relative error measurement should be applied instead.
Again, considering the large differences in the magnitude of our data, we are in favor of choosing a 
L1 loss function over a L2 loss function, as it is more robust regarding outliers. 
Hence, we decided to use the Mean Absolute Percentage Error (MAPE, cf. Equation~\ref{eq:mape}) as the scoring function, where $y_i$ are true values and $\hat{y}_i$ are predicted values, $n$ is the total number of samples.

\begin{equation}
\text{MAPE} = \frac{1}{n} \sum_{i=1}^{n} \frac{|y_i - \hat{y}_i|}{y_i} 
\label{eq:mape}
\end{equation}


\subsection{Portable Code Features}

Our approach for execution time and power prediction makes use of 
portable code features which are independent of GPU platforms. 
In other words, the code features are reused for other GPUs once they are recorded.
Therefore, creating a new prediction model for another GPU only requires to record the target values, making our approach lightweight and portable. 
Disadvantages are missing information like cache hit rate or register spilling. 

Thus, features must not depend on the GPU used, leaving a choice of possible features which are mainly covered by 
instruction counts. Since the kernel launch configuration, which is grid/block size of a kernel and size of shared memory allocation, has significant impact on kernel execution, it is also used as feature by the model.
Because these features do not change across different GPUs, only the target values have to be measured again.

\subsection{Feature Acquisition and Engineering}
\label{sec:features}

Instruction counts can be measured on different levels of abstraction: for NVIDIA GPUs, SASS \cite{StephensonFlexiblesoftwareprofiling2015} and PTX instruction sets are 
viable candidates. 
Since our approach aims for portable features which do not depend on the 
hardware, PTX seems to be the better fit as it is portable across different GPU architectures.
Usually, nvprof would be the natural choice to profile kernels, but as it 
profiles on SASS level it does not provide the required portability. 

Instead, we use the CUDA Flux profiler to gather features at PTX level \cite{cudaflux}. 
This profiler analyses the code for PTX instruction statistics on basic block level \cite{GruneModernCompilerDesign2012}, and uses code instrumentation to keep track of how often threads execute a specific basic block.
Notably, each thread of a given kernel launch is instrumented.
Besides the instruction counts, the kernel launch configuration is recorded,
 including grid and block size of the kernel and shared memory usage. 
To keep the instrumentation lightweight, only the basic block execution frequencies and PTX instruction counts for each basic block are recorded when an instrumented application is executed. 

%
%




CUDA Flux allows gathering instruction counts for each possible 
PTX instruction including specializations, thus possibly 
hundreds of features. 
These fine-grained features may contain a lot of information on the 
application, but more features also mean that the average importance for each 
feature in the model can be quite low. Moreover, a high-dimensional feature space also requires
a large amount of training data to perform the prediction. 
For the sake of a simple and comprehensible model the different instruction types are formed into more general groups of instructions.
Our experiments showed that out of the many possibilities of grouping strategies, simple grouping by arithmetic, special, logic, and control flow already yields a reasonable performance. This choice was inspired by the 
classification of PTX instructions by Patterson and Hennessy 
\cite{PattersonComputerOrganizationDesign2012}.
Regarding Patterson's proposal, the bit width of computations is ignored in order to reduce the number of features.
Moreover, memory instructions are also grouped differently to reduce the number of features. 
On the contrary, memory instructions are grouped differently, because we think that in this case the width of a memory access makes a significant difference.

For memory instructions the most important metric is the data volume which is read or written, as well as 
the memory type being used. 
Hence, the memory instructions are used to 
compute data transfer volumes for memory types 
including global memory, shared memory and parameter memory. 
Where parameters are stored depends on the implementation, but usually it is either register space or global memory.
Note that register spilling cannot be accounted for as this behavior is device dependent.
In addition to the count of each instruction group, the ratio of arithmetic 
instructions and data transfer volume of global and local memory is computed and used as input feature.

\subsection{Model Construction and Training Procedure}

The model is constructed using the Extremely Randomized Trees 
Regression method provided by the scikit-learn library \cite{scikit-learn}.
Compared to the currently pervasive interest in neural networks, the random forest methods
require a smaller amount of samples and need less training time. 

Training of a model includes a search of optimal hyperparameters, which is commonly defined by cross-validation. 
Using simple cross-validation is in general more biased than advanced methods as proposed by Cawley and Talbot \cite{CawleyOverfittingModelSelection2010} or Tibshirani \cite{TibshiraniBiasCorrectionMinimum2009}.
For our problem, the nested cross-validation is then considered because several iterations ensure good generalization. 
In each iteration a different random 
initializer is used for the splits of test and training data. First the scores of each hyperparameter combination are computed on all splits, then the best parameter combination is used to compute scores on all splits again. 

Since random forest can only learn values in the range of the training samples, we employed our own custom split for time prediction, which always includes the five samples with the longest execution time in the training set 
in order to ensure sufficient coverage of the prediction interval.
Furthermore, the custom split ensures that each split has about the same amount of samples for short (t<1,000us), medium (1,000us<=t<100,000us) and long-running (t>100,000us) kernels.
Note that this methodology requires significantly more computational resources when using more samples for training. 
If the training time is an issue, the Tibshirani method \cite{TibshiraniBiasCorrectionMinimum2009} with only two cross-validations might be a suitable alternative. 

One of main parameters for adjusting the Extremely Randomized Trees Regression is the number of estimators, which represents the number of trees in the forest. In general, the more trees are used, the better prediction quality is. However, as a large number of trees can also lead to overfitting for noisy data \cite{Segal2004a}, this parameter should not be chosen to be arbitrary large. 
Our preliminary experiments showed that using more than 1024 estimators is more likely to lead to overfitting. Therefore, in order to minimize the parameter space and reduce training time, the following parameters are used in our nested cross-validation:\\
\begin{itemize}
	\item \textbf{Max features:} max, log2 or sqrt.
	\item \textbf{Split criterion:} MSE or MAE.
	\item \textbf{N estimators:} 128, 256, 512 or 1024.
\end{itemize}

Moreover, the maximum features method that shall be used to determine best split was also added to the hyperparameters,
and the criterion for computing the quality of the split was used with MSE or MAE.

%

%% file: 04-groundtruth.tex
\section{Ground Truth}

\subsection{Benchmarks}
\label{sec:benchmarks}

To maximize the number of samples for training and evaluation of the model, we tried to use as many workloads as possible.
The used benchmark suites include: 
Rodinia 3.1 \cite{CheRodiniabenchmarksuite2009},
Parboil 2.5 \cite{StrattonParboilRevisedBenchmark},
SHOC \cite{DanalisScalableHeterogeneousComputing2010} and
Polybench-GPU 1.0 \cite{Grauer-GrayAutotuninghighlevellanguage2012}.

Adhinarayanan et al. \cite{Adhinarayananautomatedframeworkcharacterizing2016} 
characterized the benchmark suites SHOC, Parboil and Rodinia, and found that all 
benchmark suites have some unique applications. Even though some may be 
slightly over-represented, we decided to include all usable applications. 

Due to limitations of the LLVM compiler framework the CUDA Flux profiler is 
built upon, benchmarks using texture memory cannot be considered.
Table~\ref{tab:benchmarks} lists all applications and whether they are included in this analysis. 
Some applications could not be used for the power prediction.
The included benchmarks also include some irregular workloads.
For the excluded applications, the reasons are reported as well.

\begin{table}[!ht]
	\begin{threeparttable}
	\small
	\renewcommand{\arraystretch}{0.85}
	\begin{tabular}{l|l|l|l|l}
		\textbf{Suite} & parboil-2.5   & polybench-gpu-1.0 & rodinia-3.1             & shoc                  \\ \hline
		\multirow{15}{*}{\textbf{Included}} & cutcp                 & 2DConvolution        & 3D                & BFS (irregular)\\
		& histo         & 2mm               & b+tree (irregular)      & FFT                   \\
		& lbm           & 3DConvolution     & bfs (irregular)         & MD5Hash               \\
		& mri-q         & 3mm               & backprop                & MaxFlops              \\
		& sgemm         & atax              & dwt2d                   & Reduction             \\
		& stencil       & bicg              & euler3d                 & S3D                   \\
		& tpacf         & correlation       & gaussian                & Scan                  \\
		&               & covariance        & heartwall               & Sort                  \\
		&               & fdtd2d            & lud$\_$cuda             & Stencil2D             \\
		&               & gemm              & myocyte                 & Triad                 \\
		&               & gesummv           & needle                  &                       \\
		&               & gramschmidt       & particlefilter$\_$naive &                       \\
		&               & mvt               & particlefilter$\_$float &                       \\
		&               & syr2k             & sc$\_$gpu               &                       \\
		&               & syrk              &                         &                       \\ \hline
		\multirow{11}{*}{\textbf{Excluded}} & bfs\tnote{1}          & correlation (power pred.) & b+tree (power pred.)   & FFT (power pred.)  \\
		& mri-gridding\tnote{2} &                      & gaussian (power pred.) & GEMM\tnote{5} \\
		& sad\tnote{1}  &                   & hotspot\tnote{2}        & MD\tnote{1}           \\
		& spmv\tnote{6} &                   & hybridsort\tnote{1}     & MaxFlops (power pred.)     \\
		&               &                   & kmeans\tnote{1}         & NeuralNet\tnote{4}    \\
		&               &                   & leukocyte\tnote{1}      & QTC\tnote{1}          \\
		&               &                   & mummergpu\tnote{1}      & Sort (power pred.)         \\
		&               &                   & nn\tnote{2}             & deviceMemory\tnote{1} \\
		&               &                   & pathfinder\tnote{2}     & spmv\tnote{1}         \\
		&               &                   & srad-v1\tnote{6}        &                       \\
		&               &                   & srad-v2\tnote{6}        &                      
	\end{tabular}
	\begin{tablenotes}
		\scriptsize
		Exclusion reasons: \tnote{1}texture memory, \tnote{2}CUDA Flux compilation error, kernel is not loopable, \tnote{4}hardcoded datasets,	\tnote{5}no instrumentation possible due to cuBlas use, \tnote{6}unstable behavior   		
	\end{tablenotes}
	\renewcommand{\arraystretch}{1.0}
	\end{threeparttable}
	\caption{\label{tab:benchmarks}List of included and excluded applications. Some kernels are only excluded for the power prediction model.}
\end{table}

As the Polybench-GPU benchmark suite has hard-coded problem sizes, we 
decided to modify these benchmarks to allow for larger problem sizes. A longer 
execution time of a kernel is especially helpful for more accurate power readings. 
As \cite{Johnston_2018} is using four problem sizes for generating 
the features, we followed this approach.
Further modifications were also implemented when kernel and kernel call are 
not in the same compilation module, as this is not supported by the CUDA Flux 
profiler.

Power measurements are in particular sensitive to short-running kernels, as the sampling frequency is limited.
To obtain representative power values for short kernels, we therefore inserted for-loops. 
However, 
as kernels might have data dependencies, repeated executions potentially can change execution behavior.
Thus, we exclude kernels showing different output results before and after inserting for-loops.

\subsection{Data Acquisition}

Statistical data for execution time and power consumption for the GPU kernels 
of the four benchmark suites are gathered on five different NVIDIA GPUs (see Table~\ref{tab:gpus}).
Clocks for all GPUs are fixed at the shown frequency in the table with exception of the GTX1650 GPU, which is a consumer device and does not support a fixed frequency. For this device the frequency range is listed instead.
At the time of collecting the metrics the used CUDA SDK version was 9.2 and the LLVM version was 7.0.

\begin{table}[!ht]
\centering
\small
\begin{tabular}{l|l|r|r|r|r|r|r|r|r}
   	     &      	& float perf. 	& Mem. BW 	&  		& CUDA 	& Core Clock 		& Mem. Clock 	& TDP & $f_s$ \\
GPU		 & Class   	& [TFLOP/s] 	& [GB/s]	& SMs	& Cores	& [MHz]				& [MHz]        	& [W] & [Hz]  \\ 
\hline
K20      & Kepler 	& 3.5       	& 208   	& 13	& 2496	& 706        		& 2600         	& 225 & 73.6  \\ 
Titan Xp & Pascal 	& 12.0      	& 548   	& 30	& 3840	& 1404       		& 5705         	& 250 & 60.2  \\ 
P100     & Pascal 	& 9.3       	& 732   	& 56	& 3584	& 1189   & 715          	& 300 & 61.1  \\ 
V100     & Volta  	& 14.0      	& 900   	& 80	& 5120	& 1290   & 877          	& 300 & 61.2  \\ 
GTX1650 & Turing 	& 3.0       	& 128   	& 14	& 896	& 300 - 2250       	& 400 - 4001 	& 75  & 10.9  \\ 
\end{tabular}
\caption{Overview of used GPUs and their relevant hardware specifications. $f_s$ stands for power sampling frequency.}
\label{tab:gpus}
\end{table}


\subsubsection{Execution Time}

Time measurements are repeated ten times to decrease the probability of outliers. 
For each combination of benchmark and dataset all kernel executions are recorded. 
With the benchmark name, dataset and the launch sequence the time measurements can be joined with the features, which the CUDA Flux profiler provides. 
Note that some workloads execute kernels multiple times with the same parameters, thus
only the median of these time measurements is used to create a sample.
Grouping identical kernel executions reduces the number of samples from over 900,000 to about 21,000.

\begin{figure}[!ht]
	\includegraphics[width=\linewidth]{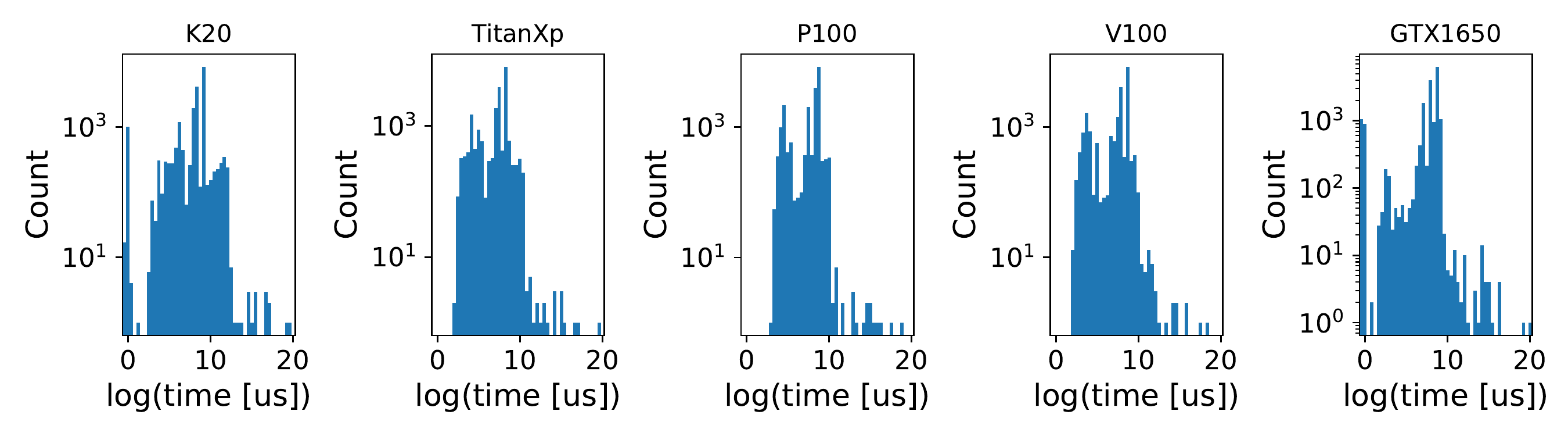}	
	\caption{Histogram of the kernel execution time in 
		logarithmic time scale. Note that long-running kernels are statistically under-represented.}
	\Description{Histogram of the kernel execution time in 
		logarithmic time scale. Note that long-running kernels are statistically under-represented.}
	\label{fig:RT_Dist}
\end{figure}

Kernel launches of the same kernel with different arguments are not grouped. 
The vast majority of samples has an execution time of less than a few 
tenths of a second (Figure~\ref{fig:RT_Dist}). 
Using GPUs with higher operating frequency or more processing units reduces the execution time even further. 
As one can see, the kernels running longer than a few seconds are under-represented. 
Because the range of kernel execution time is very large, we decided to apply the log function before training the model. 
Thus, the data is more equally distributed in the mapped space, and prediction quality improves accordingly.

\begin{figure}[!ht]
	\includegraphics[width=0.75\linewidth]{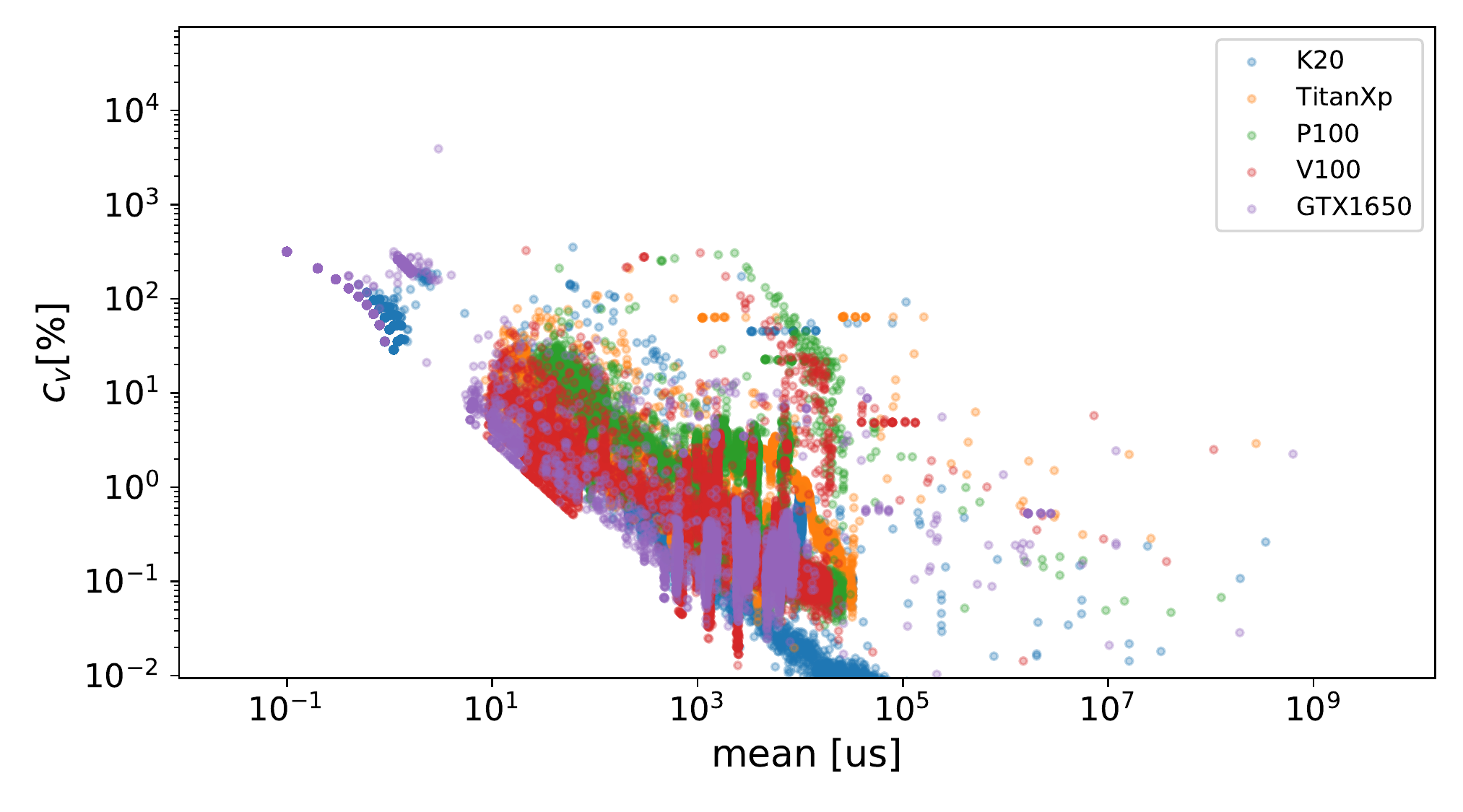}	
	\caption{Visualization of the variance of execution time: coefficient of variation $C_v$ plotted over the median of execution time (for identical kernel executions) shows that short-running kernels appear to have a larger variance compared to long-running kernels.}
	\Description{Visualization of the variance of execution time: coefficient of variation $C_v$ plotted over the median of execution time (for identical kernel executions) shows that short-running kernels appear to have a larger variance compared to long-running kernels.}
	\label{fig:RT_deviation}
\end{figure}

For very short-running kernels, for instance 1 ms and less, we expect the execution time to vary substantially. 
This has potentially also a negative impact on the prediction accuracy. 
Figure~\ref{fig:RT_deviation} shows the coefficient of variation over execution time, and demonstrates the argument above.
Furthermore, one can see that for kernels running longer than 1 ms, the coefficient of variation is reasonably low.
Still, since there is a number of measurements with a high coefficient of variation, more measurements can be beneficial for the statistical soundness of the data.

\subsubsection{Power Consumption}

Comprehensive power instrumentation and measurement are still a tedious task, 
mainly due to the lack of a complete monitoring environment for all possible 
power consumers within a given computing system.
However, for certain components of such systems, some vendors, including 
NVIDIA and Intel, provide power measurement support.
For instance, NVIDIA GPUs can be instrumented using \textit{nvidia-smi} \cite{noauthor_nvidia_2012}.
Still, the details about its functionality are poorly documented, in 
particularly, how current and voltage are measured.
Other alternatives usually require hardware access to the system, and are based 
on interposers that possibly degrade physical properties of other connections, 
including high-speed serial transmissions.
Thus, power measurement based on vendor tools is typically accepted by the 
community.
For the on-board power sensor of K20 GPUs, a detailed analysis has been performed, an error of 5\% on the order of ten power samples, while the sampling 
frequency of the sensor is approximately 66.7Hz \cite{Burtscher2014MeasuringGP}. 
In our experiments, we have been able to mainly reproduce these results, while we 
also observed that different GPU architectures and drivers result in different 
behavior regarding sampling frequency $f_s$, as shown in Table~\ref{tab:gpus}.


For power measurement, kernels are executed 
in a loop lasting at least one second, while a CPU thread records power consumption. The loop is necessary as most of the kernels have an execution time that is shorter than the measurement resolution (see Figure~\ref{fig:RT_Dist} for execution times and Table~\ref{tab:gpus} for power sampling frequency).
Multiple measurements are afterwards averaged for each kernel. 
A similar methodology can be found in \cite{5598315,8327055}. 
Similar to time measurements, the common launch sequence was used for joining the power measurements with profiling results. 

\begin{figure}[!ht]
	\includegraphics[width=0.75\linewidth]{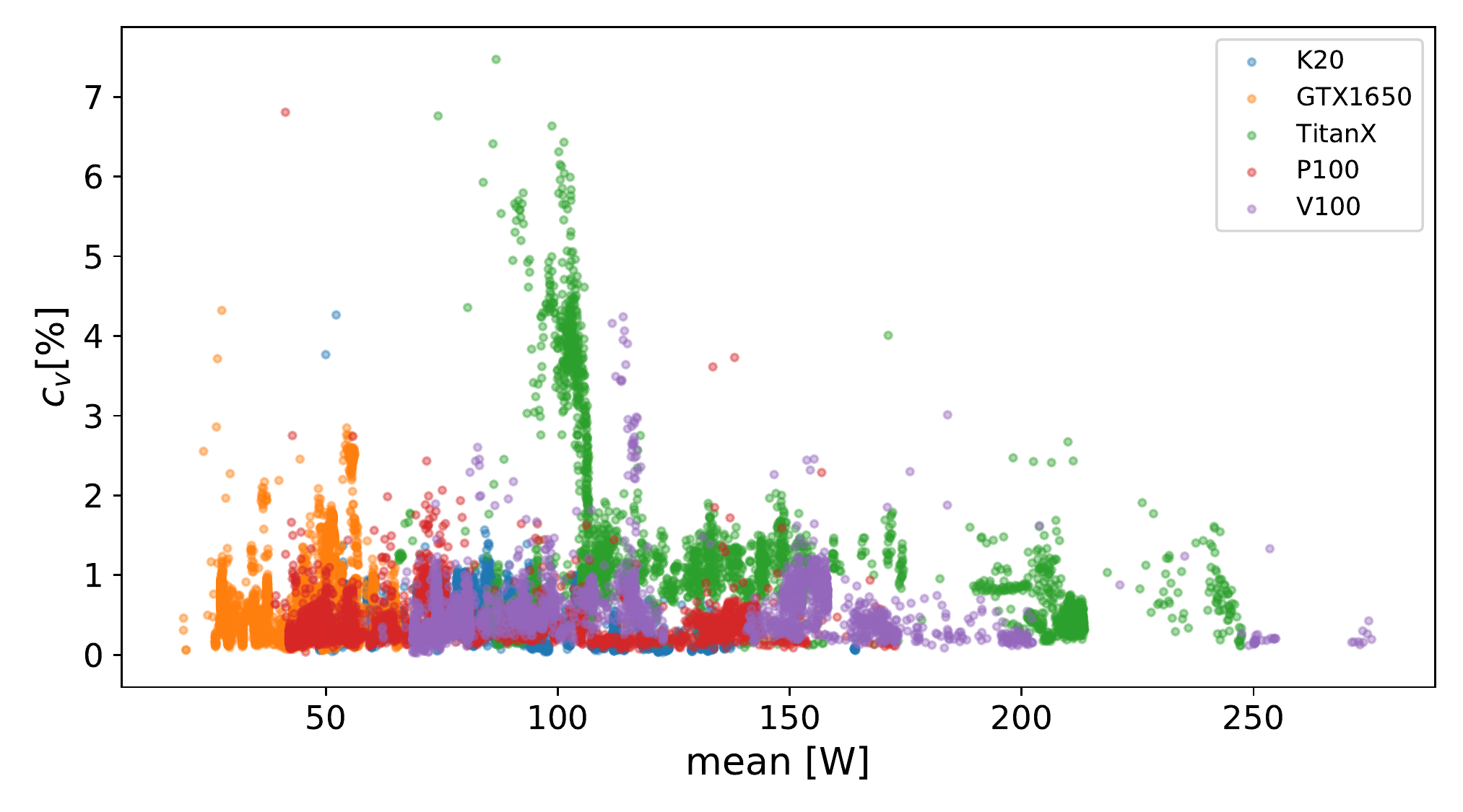}	
	\caption{Validation of power measurements by comparing the coefficient of variation against mean power consumption.}
	\Description{Validation of power measurements by comparing the coefficient of variation against mean power consumption.}
	\label{fig:RP_cv_mean}
\end{figure}

For the sake of reliability, the power measurements are repeated ten times in order to obtain representative data. 
In Figure~\ref{fig:RP_cv_mean}, the coefficient of variation versus mean value is reported.
This shows 
that the coefficient of variation of power measurements is about less than 5\%, similar to results reported in \cite{Burtscher2014MeasuringGP}.

\subsubsection{Reduction of Over-Represented Kernels}

Some kernels are executed in loops with slightly changed launch configuration or parameters.
This leads to an over-representation of some kernels which have 
thousands of samples. 
To address this, we decided to implement a threshold for the number of samples per combination of application, problem size and kernel during the random selection process.
When the threshold is too large, the kernel over-representation cannot be solved, while if the threshold is too small, too few samples are used for training.
In our study, we decided to use a threshold of 100 samples to be randomly selected for each combination, which seems to be a good compromise for both arguments above.

%% file: 06-modelling.tex
\section{K20 Case Study}

This section will review the experimental results for execution time and power prediction for the K20 GPU.
We choose the K20 for its lower compute power and therefore better distribution of execution time measurements.
The experiments on other GPUs and the results regarding portability will be covered in the following section.

To ensure good predictions the scores of multiple nested cross-validation iterations are evaluated. 
Furthermore, we employed the leave-one-out (LOO) technique to gather comparable predictions for each sample.
LOO is a special case of K-fold cross-validation where the number of folds is equal to the number of samples.
This allows spotting outliers which are not covered well by the model.

\subsection{Execution Time Prediction}

\begin{figure}[!ht]
	\includegraphics[width=0.75\linewidth]{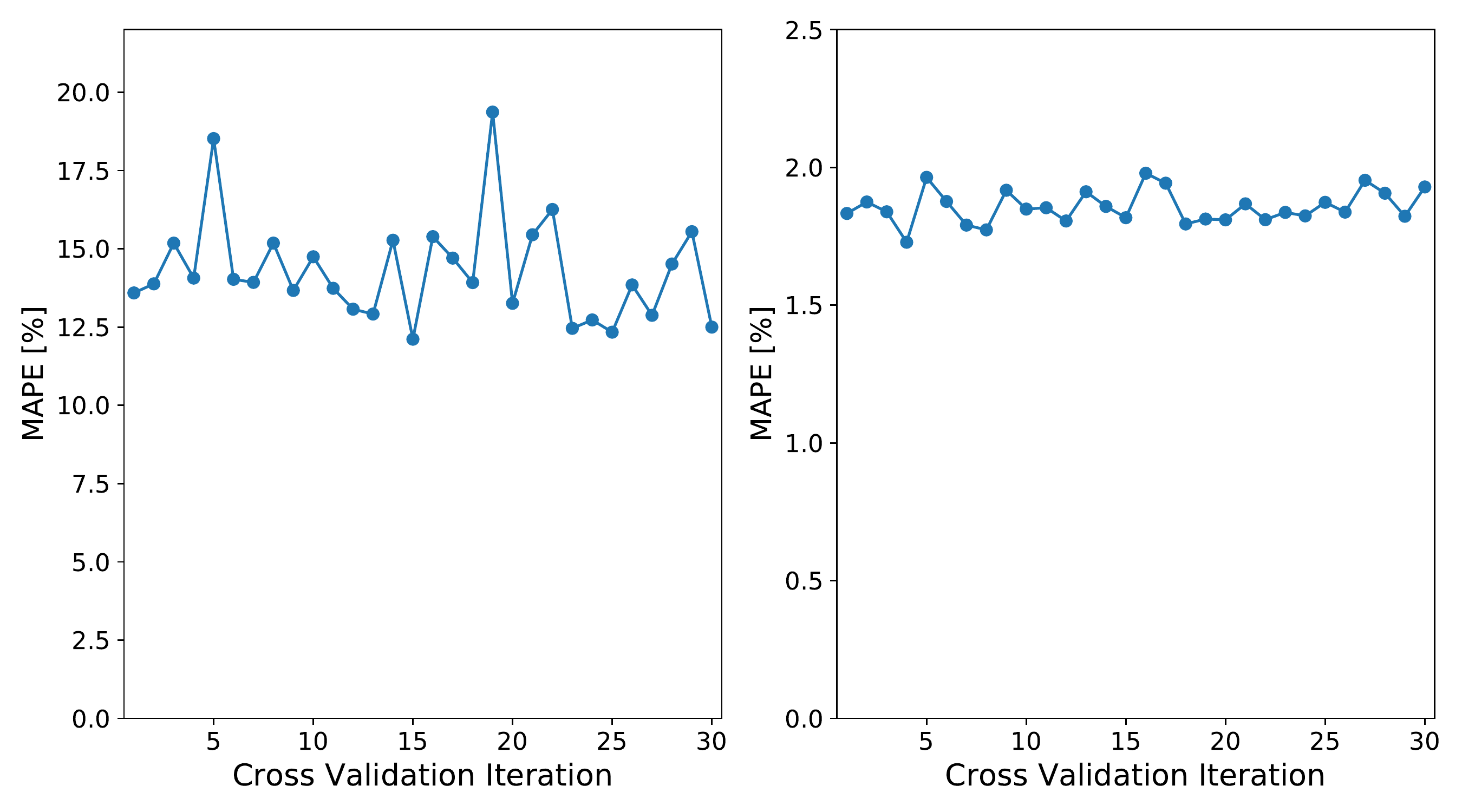}	
	\caption{Nested cross-validation score for execution time (left) and power (right) prediction on the K20 GPU.}
	\Description{Nested cross-validation score for execution time (left) and power (right) prediction on the K20 GPU.}
	\label{fig:K20_CV_Score}
\end{figure}

Figure~\ref{fig:K20_CV_Score} shows the performance of the nested 
cross-validation for time prediction. The cross-validation was 
repeated over 30 iterations with different random splits for each fold. 
Consistent and low scores indicate that the prediction generalizes well. 
The mean error (MAPE according to Equation~\ref{eq:mape}) of each iteration is in
between 12.11\% and 19.37\%. 
As different iterations show similar performance, we conclude that the prediction for the K20 can perform well with only a subset of all the samples. 

\begin{figure}[!ht]
	\includegraphics[width=0.75\linewidth]{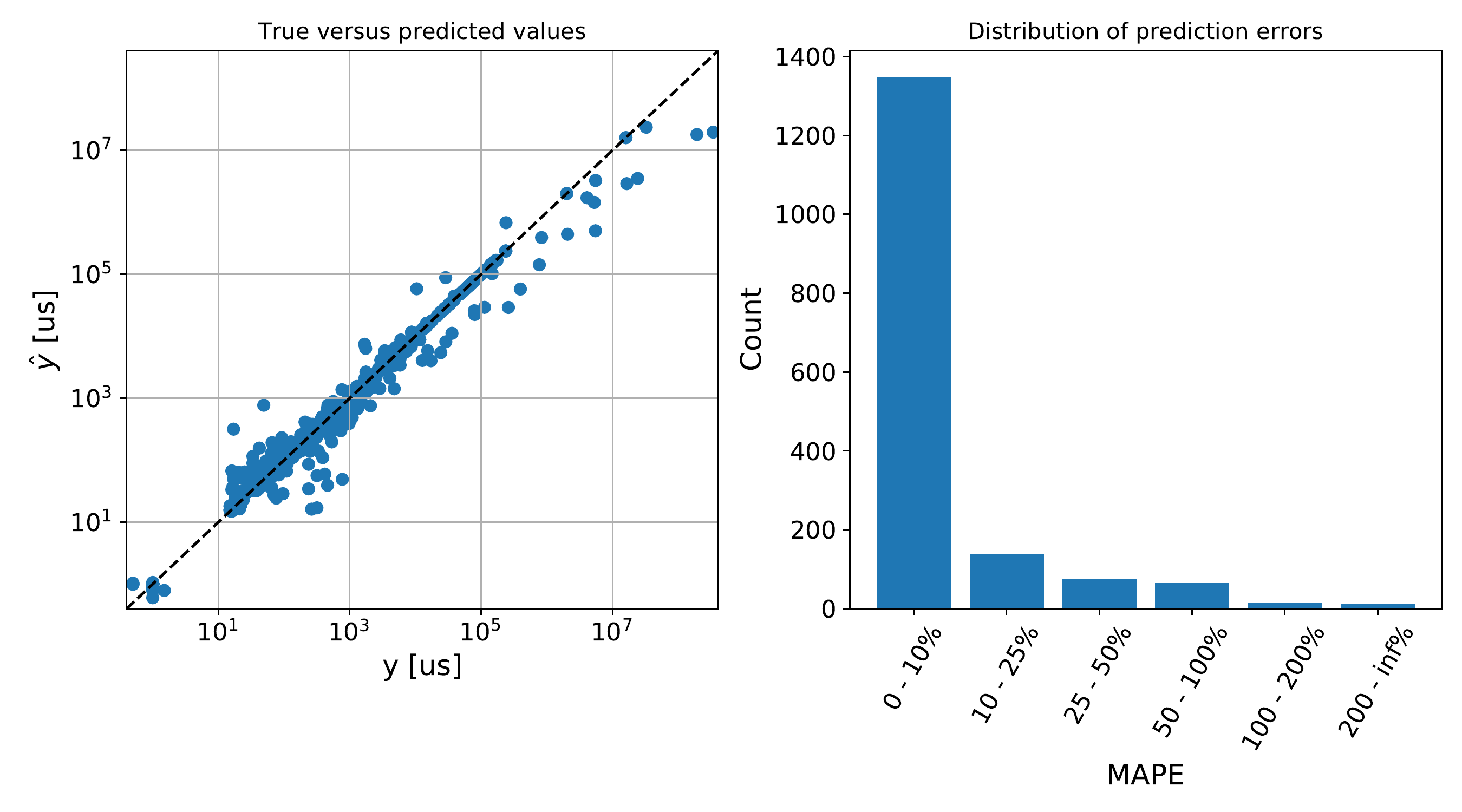}	
	\caption{Leave-One-Out results for time prediction on the K20 GPU.
		Left: scatter plot of true values versus predicted values (logarithmic scale).
		Right: distribution of prediction errors.}
	\Description{Leave-One-Out results for time prediction on the K20 GPU.
		Left: scatter plot of true values versus predicted values (logarithmic scale).
		Right: distribution of prediction errors.}
	\label{fig:K20_LOO}
\end{figure}

LOO is used to find and visualize samples which cannot be predicted well, because they are possibly outliers. 
The best parameters from nested cross-validation are used to compute predictions for each sample using the LOO method.
This method allows obtaining predictions for each sample while excluding it 
from training. 

Figure~\ref{fig:K20_LOO} shows that most of the LOO predictions are quite close to the true value. 
The samples on the high end of the prediction are usually underestimated. 
This is because random forest algorithms cannot predict values outside the range of the training samples, 
and there are only very few samples with a long execution time.
About 82\% of the samples are within 0 and 10\% of the true value,
while about 8\% are in between 10\% and 25\%. The next two groups represent both about 
4\% of the total samples. 
Only about 2\% of the samples have a deviation of more than 100\%. 
This shows that the majority of samples can predicted very well, while there are still some outliers for which the predictions deviate by a large factor.

\subsection{Power Prediction}

In this section, we follow the same methodology as for time prediction, starting with nested cross-validation score as reported in Figure~\ref{fig:K20_CV_Score}:
an error (MAPE according to Equation~\ref{eq:mape}) of in between  1.72\% and 1.97\% can be observed, effectively lower than in execution time prediction, which means that the prediction can generalize even better. 
This improvement is possibly due to the smaller range of power measurements, which have only two orders of magnitude, while for execution time measurements the range can cover up to eight orders of magnitude. 
Therefore, even few but high magnitude errors could degrade the overall performance of the nested cross-validation for time prediction, while being less likely for power prediction.

\begin{figure}[!ht]
	\includegraphics[width=0.75\linewidth]{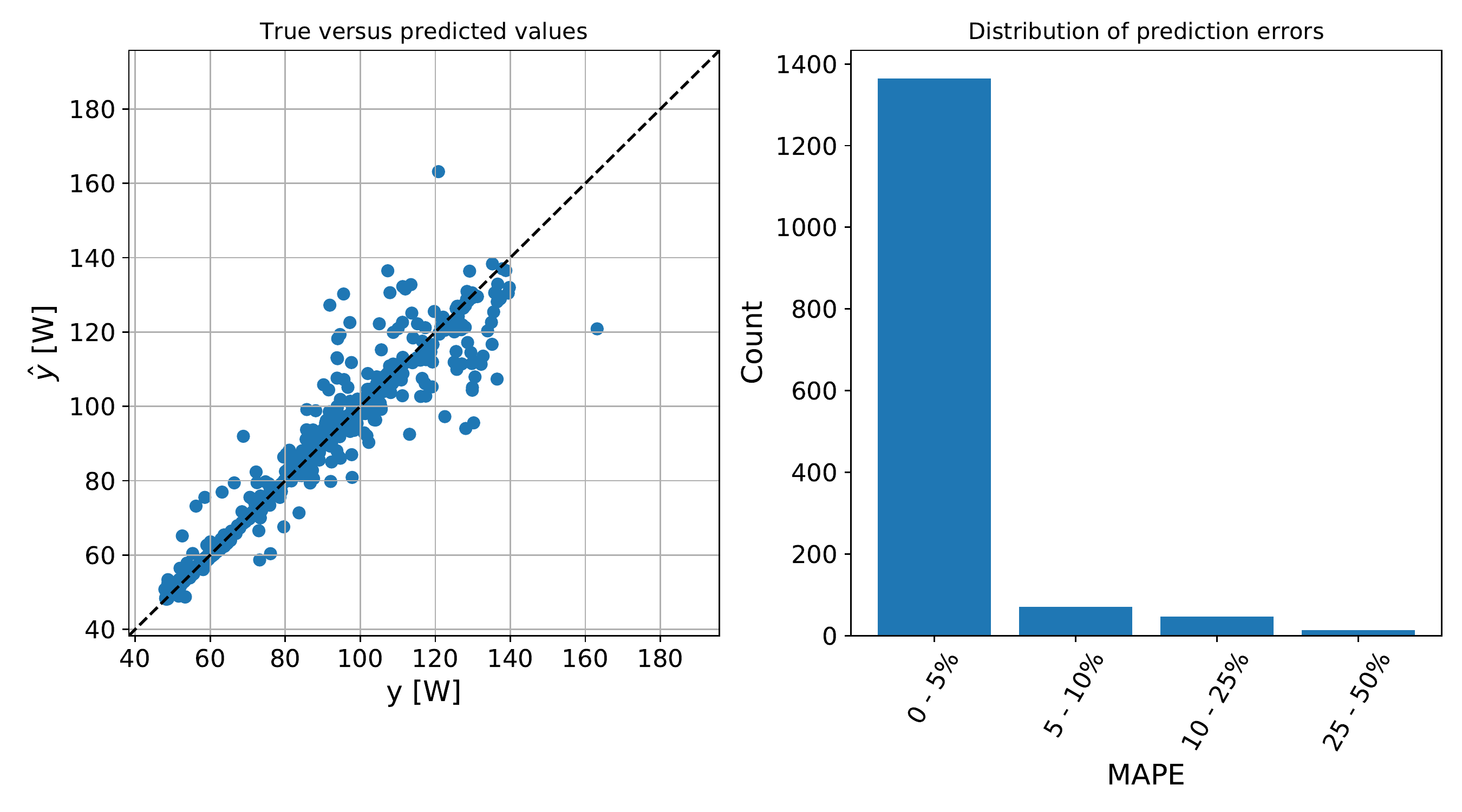}	
	\caption{Leave-One-Out results for power prediction on the K20 GPU.
	Left: scatter plot of true values versus predicted values (note the linear scale).
	Right: distribution of prediction errors.}
	\Description{Leave-One-Out results for power prediction on the K20 GPU.
	Left: scatter plot of true values versus predicted values (note the linear scale).
	Right: distribution of prediction errors.}
	\label{fig:K20_LOO_power}
\end{figure}

Last, again the LOO method is used to find possible outliers.
Using this method, the true versus the predicted values and the distribution of the prediction error are plotted in the left respectively right part of Figure~\ref{fig:K20_LOO_power}.
Most of the predictions are quite close to the true value, with 92\% of the samples being within 0 to 5\% of the true value. Only 4\% of the predictions exceed the 10\% error margin.

%% file: 07-portability.tex
\section{Portability}

This section discusses the portability of the concept, by evaluating prediction quality for all five GPUs.
As stated in the methodology, we collect application statistics (input features) only once, while for each GPU a separate output is measured (ground truth).

\subsection{Time Prediction}

\begin{figure}[!ht]
	\includegraphics[width=0.75\linewidth]{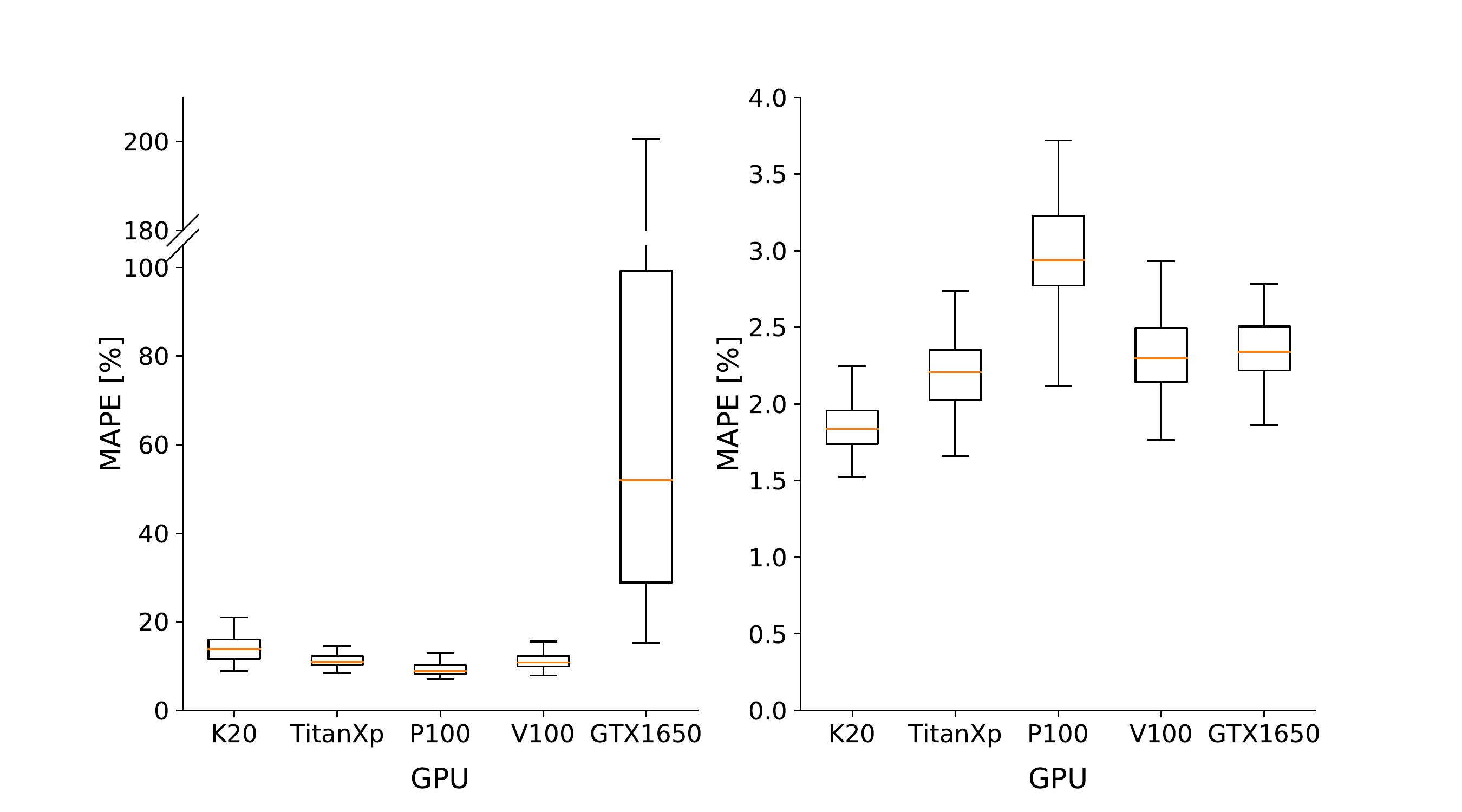}	
	\caption{Portability of time (left) and power (right) prediction across different GPUs: MAPE scores for all iterations of nested cross-validation with 
	median, first and third quartile. Whiskers are limited to 1.5 times of the 
	interquartile range (Q3-Q1). Outliers are not shown.}
	\Description{Portability of time (left) and power (right) prediction across different GPUs: MAPE scores for all iterations of nested cross-validation with 
	median, first and third quartile. Whiskers are limited to 1.5 times of the 
	interquartile range (Q3-Q1). Outliers are not shown.}
	\label{fig:SC_Scores}
\end{figure}

The results of nested cross-validation across all five GPUs are summarized in Figure~\ref{fig:SC_Scores}. 
We decided to use a boxplot of the individual scores of the folds rather than the mean score of each iteration. This avoids smoothing scores of folds with poor performance by averaging the score with possibly much better performing folds.
The median MAPE score is ranging from 8.86\% to 13.86\% for the K20, Titan Xp, P100 and V100, while for the GTX1650 it is about 52\%.
In this regard, we observe that server-class GPUs have a better predictability compared to consumer-class GPUs.
This is not surprising as the GTX1650 does not support a fixed core and memory frequency (Table~\ref{tab:gpus}).

We furthermore observe that the GTX1650 has a much higher variability compared to the other GPUs:
while the median MAPE score of 52\% is already quite high, the third 
quartile of 99.23\% leads to a large interquartile range 
(IQR), indicating a high variability and therefore poor generalization of the 
model. 

\begin{figure}[!ht]
	\includegraphics[width=\linewidth]{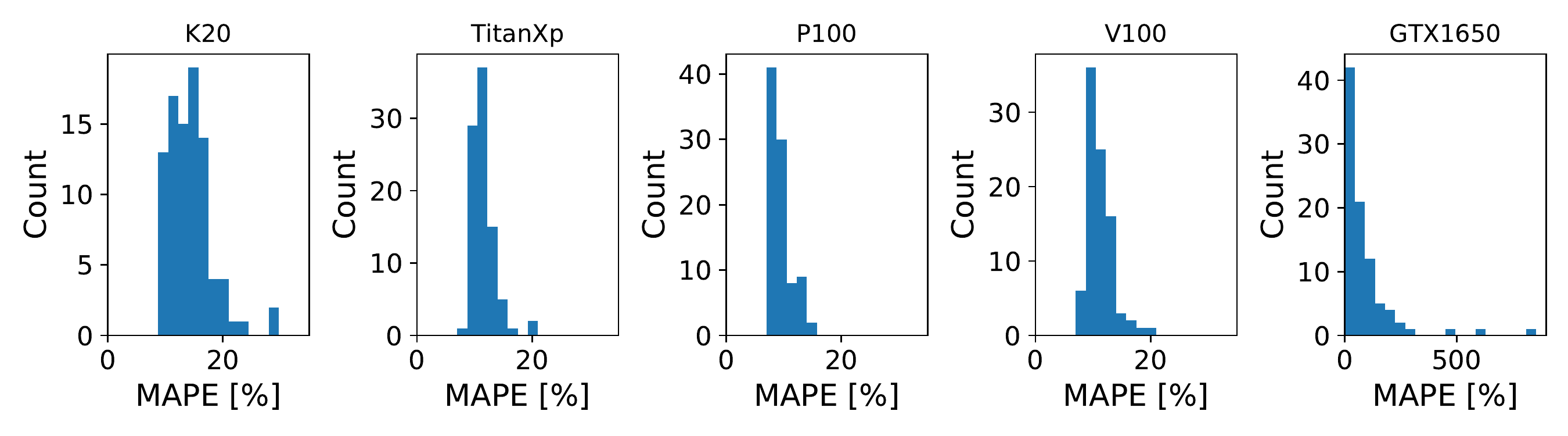}	
	\caption{Histogram of the MAPE score for each fold of the 
		nested cross-validation.}
	\Description{Histogram of the MAPE score for each fold of the 
		nested cross-validation.}
	\label{fig:CV_Histo_time}
\end{figure}

First, we analyze the distribution of errors for the different GPUs.
Figure~\ref{fig:CV_Histo_time} shows the histograms of MAPE scores of each GPU for all cross-validation iterations.
The scores for the GTX1650 are especially widely spread, but there is still a reasonable amount of samples with low error scores. 
This suggests that the dataset may contain outliers or at least very unique 
samples which are hard to predict if as there are no similar samples in the 
training set. 
This could be due to the dynamic frequencies of the GTX1650 and also due to lower overall accuracy measuring time on this device.


\begin{figure}[!ht]
	\includegraphics[width=\linewidth]{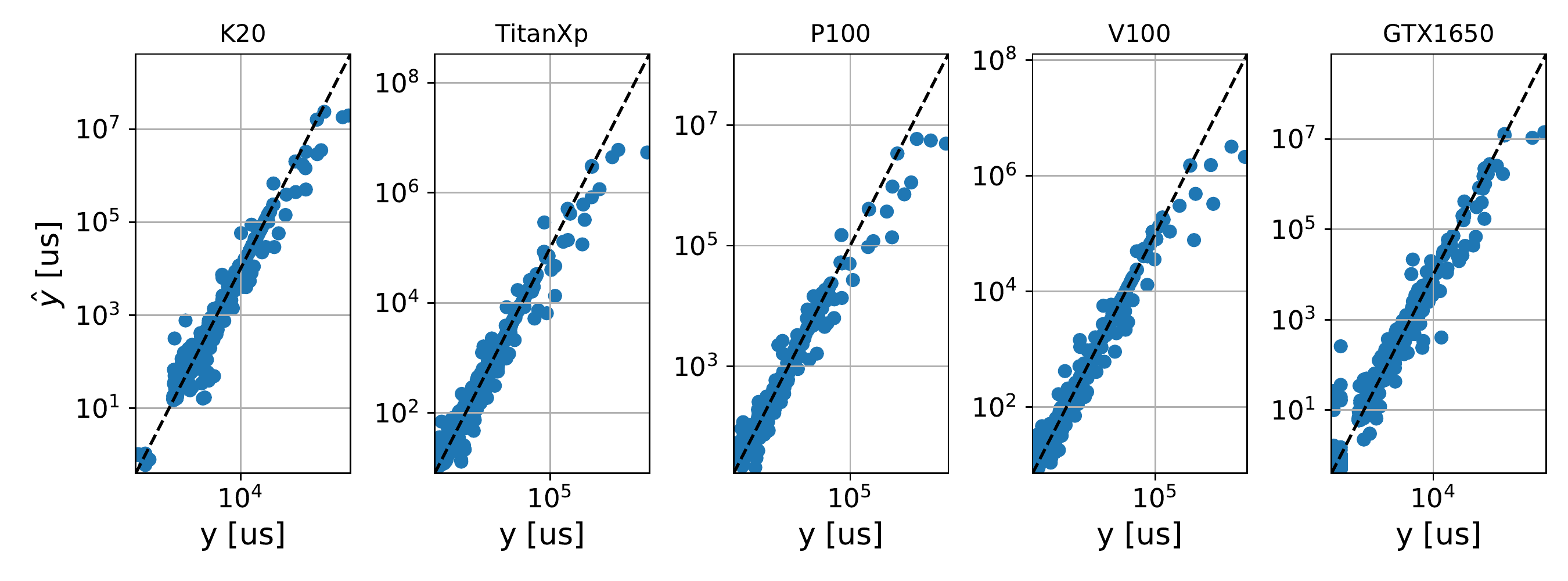}	
	\caption{Scatter plot of true time values versus predicted values 
		using the leave-one-out method for K20, GTX1650, Titan Xp, P100 and V100 
		GPUs.}
	\Description{Scatter plot of true time values versus predicted values 
		using the leave-one-out method for K20, GTX1650, Titan Xp, P100 and V100 
		GPUs.}
	\label{fig:LOO_Scores}
\end{figure}

As the CUDA drivers can typically add about 1-50~us of latency to a kernel execution time, 
depending on configuration and iterativeness, measurements of short kernels can become unreliable.
As a result, it is harder to fit such measurements into a model. 
We use again the Leave-One-Out method to accurately assess the prediction 
performance for every single sample: 
Figure~\ref{fig:LOO_Scores} reports the corresponding scatter plots, 
in which one can see that for the GTX1650 the amount of samples with short execution time $y$ is much higher in comparison to the other GPUs. 
We also see evidence for the under-representation of long-running kernels, 
as the error increases substantially for the samples with long execution time.


\begin{table}[!ht]
	\small
	\begin{tabular}{ l | l | l | l}
		GPU      & Best hyperparameters              & Avg. depth & Prediction latency \\ 
		\hline
		K20      & MAE, max features, 512 estimators & 34.83      & 108.56 ms          \\  
		Titan Xp & MAE, max features, 512 estimators & 33.30      & 108.65 ms           \\ 
		P100     & MAE, max features, 512 estimators & 34.94      & 108.30 ms          \\ 
		V100     & MAE, max features, 512 estimators & 34.30      & 106.91 ms          \\ 
		GTX1650 & MSE, max features, 256 estimators & 31.06      & 107.46 ms          
	\end{tabular}
	\caption{Hyperparameters for the best model for time prediction, together with the corresponding
	average prediction latency, measured on a Intel Xeon E5-2667 v3 CPU.}
	\label{table:time_hyperparameters}
\end{table}

Last, we report optimal hyperparameters as the result of cross-validation runs in Table~\ref{table:time_hyperparameters}.
For performance comparison we also added the average tree depth and the prediction latency.
Note that the corresponding average prediction latency for these hyperparameter settings is constantly low, but still varies substantially with hyperparameter configuration.
This suggests that the latency could be reduced by a more sophisticated hyperparameter search. 
Prediction latency was measured on an Intel Xeon E5-2667 v3 CPU clocked at 3.2~GHz.

\subsection{Power Prediction}

The results of the nested cross-validation for power are summarized in the right side of Figure~\ref{fig:SC_Scores}. 
The median of MAPE for the K20 GPU is below 2\%, and for the other GPUs in a comparable range. 
This shows a constantly good prediction, even though the peak power consumption of the GPUs varies substantially (Table~\ref{tab:gpus}).

\begin{figure}[!ht]
	\includegraphics[width=\linewidth]{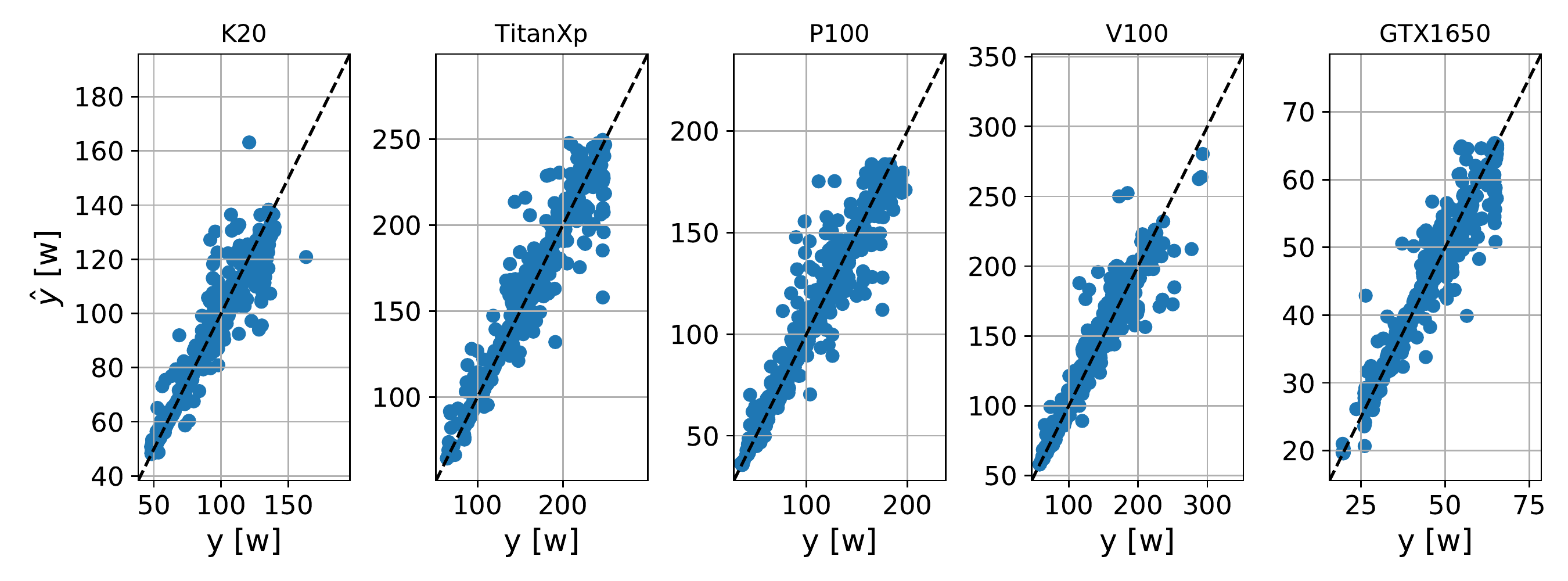}	
	\caption{Scatter plot of true power values versus predicted values 
		using the leave-one-out method for GTX1650, Titan Xp, P100 and V100 
		GPUs.}
	\Description{Scatter plot of true power values versus predicted values 
		using the leave-one-out method for GTX1650, Titan Xp, P100 and V100 
		GPUs.}
	\label{fig:LOO_Scores_power}
\end{figure}

Kernel power consumption was measured in the order of milli-Watts. 
Performing leave-one-out, we are able to plot the model predictions and actual power values as shown in Figure~\ref{fig:LOO_Scores_power}, to identify extreme outliers.
It turns out that those few outliers are different kernel samples having identical kernel features, however, exhibiting different power consumption.
Therefore, for those cases the model is not able to predict both kernels precisely. This can be attributed to limited features or potential statistical variance.

Both execution time and power prediction latency depend on the number of estimators (number of trees) and the average maximum depth of the trees.
Increasing the number of trees and their depth also increases the number of operations for traversing all trees \cite{louppe2014understanding}, which results in higher prediction latency. The latency for power is, even with a high number of trees, well below 100 ms (Table~\ref{table:power_hyperparameters}).

\begin{table}[!ht]
	\small
	\begin{tabular}{l | l | l | l}
		GPU      & Best hyperparameters              & Avg. depth & Prediction latency \\ 
		\hline
		K20      & MAE, max features, 256 estimators  & 32.08     & 15.36 ms           \\ 
		Titan Xp & MAE, max features, 256 estimators  & 33.45     & 15.32 ms           \\ 
		P100     & MAE, max features, 512 estimators  & 32.49     & 30.14 ms           \\ 
		V100     & MSE, max features, 1024 estimators & 32.91     & 60.58 ms           \\ 
		GTX1650 & MAE, max features, 1024 estimators & 32.19     & 59.20 ms           \\ 
	\end{tabular}
	\caption{Hyperparameters for power prediction model, together with the corresponding
	average prediction latency, measured on an Intel Xeon E5-2667 v3 CPU.}
	\label{table:power_hyperparameters}
\end{table}


\subsection{Feature Importance}

Notably, we observe that feature importance varies within different GPUs.
Feature importance for a fast prediction is important, as prediction time can be reduced by solely relying on a limited amount of features, albeit probably at the cost of accuracy.
Here, we will not perform such a trading, but shortly discuss which effects we observe on feature importance and how this can be explained with the particular GPU architecture.

\begin{table}[!ht]
\small
\begin{tabular}{ l | r | r | r | r | r | r | r | r | r | r }
					& \multicolumn{5}{c}{Time} 				& \multicolumn{5}{|c}{Power} 				\\
					&    	& Titan  & 			&  		& GTX	& 		& Titan & 		&  		& GTX	\\
Feature				& K20	& Xp 	 & P100		& V100 	& 1650	& K20	& Xp  	& P100	& V100 	& 1650	\\
\hline
threads per CTA		& 23.19	& 27.17	 & 26.62	& 29.62	& 23.49	& 19.74	& 24.70	& 17.91	& 14.77	& 9.52 	\\
CTAs				& 8.47	& 10.01	 & 11.74	& 10.76	& 5.51	& 20.64	& 8.81	& 16.49	& 20.26	& 19.28 \\
total instr.		& 7.90	& 7.73	 & 6.40		& 6.34	& 9.57	& 5.58	& 6.01	& 5.84	& 4.36	& 4.10 	\\
special ops			& 1.16	& 1.53	 & 1.96		& 1.46	& 0.42	& 2.37	& 8.00	& 3.35	& 1.39	& 1.07 	\\
logic ops			& 2.15	& 2.58	 & 2.38		& 2.30	& 1.34	& 3.91	& 4.32	& 3.27	& 3.87	& 10.94 \\
control ops			& 4.41	& 4.50	 & 3.75		& 3.71	& 5.51	& 3.69	& 6.11	& 2.68	& 2.36	& 2.41 	\\
arithm. ops			& 6.96	& 8.12	 & 6.75		& 7.01	& 11.62	& 6.75	& 6.46	& 6.72	& 4.84	& 5.22 	\\
sync ops			& 2.96	& 3.54	 & 4.89		& 4.71	& 2.34	& 4.97	& 4.05	& 8.72	& 5.08	& 5.84 	\\
global mem vol.		& 12.46	& 16.30	 & 16.30	& 14.13	& 15.59	& 8.28	& 10.61	& 6.47	& 5.73	& 4.99 	\\
param mem vol.		& 20.14	& 8.15	 & 9.08		& 8.60	& 13.45	& 16.63	& 11.39	& 17.72	& 27.45	& 30.27 \\
shared mem vol.		& 4.38	& 4.22	 & 4.66		& 4.97	& 4.72	& 3.88	& 3.56	& 7.49	& 7.27	& 4.77 	\\
arithm. intensity	& 5.81	& 6.14	 & 5.47		& 6.40	& 6.43	& 3.57	& 5.99	& 3.34	& 2.62	& 1.61 
\end{tabular}
\caption{Feature importance in percent for time and power prediction.}
\label{table:fi_time}
\end{table}

Table~\ref{table:fi_time} lists feature importance for time and power prediction for the different GPUs.
The order of importance of features changes across different GPUs, albeit some structure can be identified (we refer to the rank $x$ of a feature with regard to importance with \textit{\#x}).
For an overview of the GPUs' properties, please refer to Table~\ref{tab:gpus}.

The main observations and possible explanations for time predictions include that 
\textit{threads per CTA} is always of highest importance (\#1), indicating that a good utilization of an SM is important, which is inline with the requirement of parallel slackness according to the BSP model~\cite{valiant1990}.
Also, the constant importance of \textit{global mem vol.} (\#2 or \#3) indicates that a GPU's performance highly depends on memory operations, which also follows intuition as the vast amount of execution units has to be kept busy.
\textit{CTAs} becomes important with increasing SM counts (\#4 for K20, \#3 for Titan Xp, P100, and V100), except for consumer-class GTX1650, respectively also correlates with total CUDA cores.

Cumulative feature importance is similar for all GPUs with regard to number of features: 50\% of the total importance is covered considering the top 3 features. 
Note that these top 3 features can be different in order and absolute terms, as shown in Table~\ref{table:fi_time}.
However, \textit{threads per CTA} and \textit{global mem vol.}, contribute together in between 36-44\%.
Also, while the top 1 feature is similarly important across GPUs (~23-30\%), top 2 features differ more: 14.13\% for V100, while 20.14\% for K20. 
Looking at top 5 features, GTX1650 has most coverage with 73.73\%, then K20 with 72.17\%, P100 with 70.49\%, V100 with 70.12\%, and Titan Xp with 69.75\%. 
Notably, the absolute difference is rather small.

Observations and possible explanations for power predictions include
that the three most important features are \textit{threads per CTA}, \textit{CTAs}, and \textit{param mem. vol.}, except for GTX1650 and V100 (in the top 4, though). 
In detail, \textit{threads per CTA} seems most important if the overall SM count is low (\#1 or \#2 for K20, GTX1650, and Titan Xp), in particular compared to large SM counts (\#3 for P100, \#4 for V100).
Notably, \textit{param mem. vol.} becomes increasingly important if \textit{threads per CTA} importance decreases.
Overall, this suggests that utilization is a prime concern with regard to power consumption.

Importance of \textit{global mem vol.} is high if memory bandwidth is low, indicating problems keeping the SMs busy (\#3 for GTX1650 with 128 GB/s, \#4 for K20 with 208 GB/s).
Overall, it is surprising that \textit{global mem vol.} is rather unimportant (\#5 or \#7 for the other GPUs), as memory is considered to contribute substantially to overall power consumption.
Possibly, memory is not energy-proportional, resulting in a rather static power fraction, as long as the SMs are kept busy.

Similarly to time predictions, cumulative feature importance is similar for all GPUs with regard to number of features: 50\% of the total importance is covered considering the top 3 features, except for GTX1650 (46.70\%).
Top-5 feature importance is lowest for GTX1650 (63.50\%, similar to time), and similar for all other GPUs (68-78\%). Generally, cumulative feature importance tends to be highest for P100 and V100, and lowest for GTX1650 and K20.

In summary, feature importance for time predictions shows that a high utilization and keeping the SMs busy is of paramount importance. 
This is reflected by \textit{threads per CTA} being always most important, followed by \textit{global mem vol.}, and either \textit{param mem. vol.} or \textit{CTAs} (top 3).
Also, this feature set constantly contributes more than 50\% to importance, indicating that both high utilization and global memory accesses mainly determine performance.

Comparing time and power predictions,
feature importance for power is much more diverse, allowing for less reasonings on importance. 
In a direct comparison of cumulative importance, the top 7 for time consists of only 8 distinct features total (\textit{CTAs}, \textit{arithm. intensity}, \textit{arithm. ops}, \textit{control ops}, \textit{global mem vol.}, \textit{param mem vol.}, \textit{threads/CTA}, \textit{total instr.}), for a cumulative importance in between 82.36-85.67\%.
Contrary, the top 7 for power prediction consists of 11 features total (\textit{CTAs}, \textit{arithm. ops}, \textit{control ops}, \textit{global mem vol.}, \textit{logic ops}, \textit{param mem vol.}, \textit{shared mem vol.}, \textit{special ops}, \textit{sync ops}, \textit{threads/CTA}, \textit{total instr.}), for a cumulative importance in between 76.08-86.05\%.

Also, one can observe that \textit{total instr.} is consistently more important for time than for power, which seems reasonable as execution time intuitively depends on how many instructions are executed, while power consumption is rather dependent on the instruction type, which utilize different hardware units and components.
The latter intuition is also reflected in observed feature importance. 
Last, \textit{global mem vol.} is substantially more important for time than for power, which as said is possible due to a lack of energy proportionality of the memory subsystem, respectively the need to keep execution units busy.

%% file: 08-discussion.tex
\section{Discussion}
\label{sec:discussion}

The cross-validation shows that our models generalize well for predicting 
time and power. The time prediction has median MAPE results ranging from 8.86\% to 
13.86\% for professional GPUs.
The time for the consumer-class GPU GTX1650 could not be predicted as well as for server-class GPUs.
A possible reason is the dynamic core and memory frequency which cannot be made static, leading to poor measurement accuracy and making the GPU behavior hard to predict. 
The cross-validation for power prediction yields a median MAPE varying 
from 1.84\% to 2.93\% for all used GPUs. 

In spite of relying only on static input features, our results still show a very good prediction 
accuracy for five tested GPUs, both regarding time and power predictions.
With regard to portability, experiments showed that a model trained 
specifically for a given GPU can accurately predict time and power for an 
application's static set of input features.
Furthermore, an extensive use of cross-validation shows that the models 
generalize well, in spite of a rather limited dataset size.
These results suggest that learning methods such as random forests based solely on hardware-independent features can capture application behavior well to predict execution time and power consumption with sufficient accuracy.
Contrary, we observe that dynamic hardware configurations, such as varying operating frequency without possibility of control (see consumer-class GTX1650) are more difficult to be captured by a model.
A more detailed analysis of related capturing capabilities of learning methods is left for future work.

\subsection{Prediction Latency}

Predictions can be made fast, as experiments show that prediction latency
is typically in the range of 15-108 milliseconds.
While it is straight-forward to state that prediction latency should be as small as possible, concrete constraints heavily depend on the use:
for provisioning and procurement tasks, in which different system architectures are being evaluated, the process is rather bound by throughput and not latency, and can furthermore easily be parallelized.
Scheduling is usually much more diverse, as aspects such as task granularity or point of time of scheduling might differ substantially.
For load-balancing or work-stealing concepts, a sub-millisecond latency is desirable \cite{7980039}.
Contrary, for a distributed system, a scheduling latency of 10ms for a 100ms task is considered too high \cite{10.1145/2517349.2522716} (effectively 10\% of the execution time), again assuming scheduling being part of the critical path.
Contrary, for workloads with predictable behavior, scheduling can be done prior to execution, relaxing constraints substantially (possibly again in the range of 10\% of total execution time).
Also, given the offloading of GPU acceleration, scheduling the next kernel can be overlapped with current kernel execution, thereby relaxing the latency constraint to the execution time of a kernel.

We would like to add that prediction is not optimized in any way for short latency and can still be improved, and furthermore it is possible to trade accuracy for latency by using fewer trees and/or features in the model.
Second, the used benchmark suites are mostly tailored for architectural simulators, and the resulting bias to short-running kernels has been noted before.
Real applications, in particular multi-GPU ones, can have substantial larger execution time.
Thus, in particular the use for scheduling decisions, even across a variety of 
heterogeneous devices, seems feasible.

\subsection{Related Work}

As the works \cite{ArafaScalableGPU2019, GuerreiroGPUStaticModeling2019} are recent and quite similar to our study here, we shortly discuss them: 
\cite{GuerreiroGPUStaticModeling2019} is also using machine learning and PTX code of the kernels to predict the time and power consumption of CUDA kernels. 
They use recurrent neural networks, support also DVFS, and preprocess PTX code to obtain additional information on instruction dependencies.
As a result, dynamic control is not predictable and therefore loops need to be unrolled. 
Still, this trade-off allows predicting a sequence of PTX instruction, unlike this work which mostly predicts based on a histogram of the instructions.	

A hybrid approach called PPT-GPU based on an event-based simulation of GPU kernel execution in combination with an analytical model is described in \cite{ArafaScalableGPU2019}. 
This hybrid approach allows avoiding time-consuming cycle-level simulations.
Like \cite{GuerreiroGPUStaticModeling2019}, they are predicting execution time based on PTX instruction sequences and also preprocessing the PTX code for additional information like dependencies. 
Furthermore, including cache behavior in the prediction can improve the quality of results.
We tested PPT-GPU for MAPE results, using this publicly available model. 
Note that this does not yet include the cache model as described in \cite{ArafaScalableGPU2019}.
The resulting MAPE score for PPT-GPU was 433.8\%, based only on the polybench-GPU benchmark suite.
A direct comparison to our work would require a major rework of the learning methodology, including composition of training and test data set, and cross-validation.
In essence, test data set would consist only of polybench-GPU, while all other applications form the training set.
Such a major change in methodology has tremendous implications on model accuracy, thus cannot be representative.
Still, an evaluation of an according training and test procedure resulted in a MAPE of 218.6\%.
Furthermore, we would like to highlight that our prediction time is substantially shorter than the time-consuming simulation.

While predictions based on sequence of PTX instructions seem very favorable compare to histograms (our approach), 
a direct comparison to \cite{ArafaScalableGPU2019} highlights our competitiveness. 
While we are not including instruction dependencies, our model still captures conditional branches and thus supports dynamic control flow.



\subsection{Limitations}

Still, we observed a couple of limitations which we summarize in the 
following:

\emph{Training data:}
certainly a larger training data set would be helpful to improve the prediction accuracy.
Furthermore, the database of samples used to build the model mainly includes short running 
kernels. 
As measurements for kernels with short execution time are less accurate, this also limits the accuracy.
With fewer data on long-running kernel it is also harder to predict this class of samples.
Also, 14.59\% (GTX1650) to 56.08\% (V100) of all kernel launch configurations do not utilize all available streaming multiprocessors (register usage ignored), indicating that more samples with high degree of parallel work would be helpful.
Regarding power prediction, short and data-dependent kernels show unexpected behavior when for-loops are inserted for obtaining adequate power measurements, leading us to exclude them from our analysis.
A possible solution to the issues with the training data set may be synthetic workloads with 
configurable execution time and degree of parallelism, e.g. similar to the one used in \cite{ChoiRooflineModelEnergy2013}.

\emph{Model features:}
to address the increasing interest in reduced-precision arithmetics, for instance 16bit floating-point or 8bit integer, weighting the computational instructions by bit width is possible. 
In general, introducing features to reflect the degree of optimization would be helpful, respectively, indicting performance bugs like bank conflicts, branch divergence, or memory coalescing issues.
As pointed out previously, some kernels show a strong variation in between consecutive kernel launches. More research is required to understand this situation and how this can be covered by features.

\emph{Model training:}
in general, a larger hyperparameter space as well as regularization of hyperparameters could further improve prediction accuracy.
Keeping the prediction latency low while improving the accuracy may be very difficult and is at best very time-consuming if the training methods employed in this work are not optimized.


%% file: 09-summary.tex
\section{Summary}

We hypothesized that \emph{GPU kernels are usually well-structured, sufficiently optimized for locality and latency-tolerant}, therefore a prediction of execution time and power consumption based solely on hardware-independent features, which describe code and kernel launch configuration, is feasible.
We validate our hypothesis by training machine learning models for five GPUs, and evaluate their accuracy by comparing to monitored real execution of at least 184 unique kernels, using when possible different problem sizes (thus, kernel launch configurations), as ground truth.
The cross-validation shows that our models generalize well for predicting time and power. 
Median MAPE results for time prediction are 13.86\%, 10.95\%, 8.86\%, 10.89\%, 52.0\%, 
while for power prediction 1.84\%, 2.21\%, 2.94\%, 2.30\%, 2.33\%,
for a K20, Titan Xp, P100, V100 GPU and GTX1650, respectively.

We observed that the dataset, based on a representative set of benchmark suites, tends to rather short-running kernels, resulting in a poor representation of long-running kernels.
Results suggest that for consumer GPUs with dynamic core and memory frequency, like the consumer-grade GTX1650, this lack of representation amplifies, which is reflected by an increase of median MAPE error.
In contrary, the median MAPE error for power is similarly low for all GPUs. 

In summary, we conclude that our hypothesis is supported, as GPU kernel execution time and power consumption can be accurately predicted by using solely hardware-independent features.
As a result, we are proposing a portable, fast, accurate model to predict time and power consumption, which is publicly available and can be easily retrained for other GPU architectures.
Note that portability is currently limited to CUDA, which, however, is a practical and not principal limitation.

Future work can include further feature engineering and more sophisticated features describing the degree of optimization in order to improve prediction accuracy.
More effort on hyperparameter search and optimization could improve prediction latency and enhance the generalization of the models.
